%% file: main.tex
\documentclass[conference]{IEEEtran}
\usepackage{dirtree,array}
\usepackage{tikz}
\usepackage{amsmath}
\usepackage{soul, color, xcolor}
\usepackage{amsmath,amsfonts}
\usepackage{algorithmicx}
\usepackage{graphicx}
\usepackage{textcomp}
\usepackage{xcolor}
\usepackage{siunitx}
\usepackage{comment}
\usepackage{tcolorbox}
\usepackage{listings}
\usepackage{xspace}
\usepackage{cancel}
\usepackage{float}
\usepackage{graphicx}
\usepackage{subfigure}
\usepackage{pifont}
\usepackage{threeparttable,booktabs,multirow,lscape}
\usepackage{multirow}
\usepackage{adjustbox}
\usepackage{enumerate}
\usepackage{listings}
\usepackage{tabularx}
\usepackage[ruled,vlined]{algorithm2e}
\usepackage{amsmath}
\usepackage{environ}
\usepackage{tikz}
\usepackage{color}
 \let\labelindent\relax
 
 \usepackage{amssymb}
\usepackage{enumitem}
\usepackage{rotating}
\usepackage{threeparttable}
\usepackage{url}
\usepackage{hyperref}
\usepackage{bbding}
\usepackage{lastpage}
\usepackage{caption}
\input{solidity}
\ifCLASSINFOpdf
\else
\fi
\begin{document}
%
\title{Dissecting Payload-based Transaction Phishing on Ethereum}

\newcommand{\tab}{\hspace*{1em}}
\newcommand{\code}[1]{{\fontfamily{cmtt}\fontseries{m}\fontshape{n}\selectfont\small{#1}}}
\newcommand*{\prompt}[1]{\textsf{\textbf{#1}}}
\newcommand{\phish}{Semantic-based phishing scams targeting Decentralized Finance}
\newcommand{\shortphish  }{\textsc{PTXPhish}}
\newcommand{\etc}{\emph{etc}\xspace}
\newcommand{\ie}{\emph{i.e.}\xspace}
\newcommand{\eg}{\emph{e.g.}\xspace}
\newcommand{\Ie}{\emph{I.e.}\xspace}
\newcommand{\Eg}{\emph{E.g.}\xspace}
\newcommand{\legit}{normal app}
\newcommand{\leg}{normal}
\newcommand{\hypo}[1]{\textcolor{purple}{[hypo: #1]}}
\newcommand{\qy}[1]{\textcolor{purple}{[hypo: #1]}}
\newcommand{\lei}[1]{\textcolor{red}{[lwu: #1]}}
\newcommand{\sg}{UTG}


%

\author{\IEEEauthorblockN{Zhuo Chen~\thanks{~\IEEEauthorrefmark{1} Corresponding Author.}}
\IEEEauthorblockA{Zhejiang University\\
hypothesiser.hypo@zju.edu.cn}
\\
\IEEEauthorblockN{Dong Luo}
\IEEEauthorblockA{Zhejiang University\\
22321053@zju.edu.cn}
\and
\IEEEauthorblockN{Yufeng Hu}
\IEEEauthorblockA{Zhejiang University\\
yufenghu@zju.edu.cn}
\\
\IEEEauthorblockN{Lei Wu~\IEEEauthorrefmark{2}~\thanks{~\IEEEauthorrefmark{2} These authors are also affiliated at Key Laboratory of Blockchain and Cyberspace Governance of Zhejiang Province.}}
\IEEEauthorblockA{Zhejiang University\\
lei\_wu@zju.edu.cn}
\and
\IEEEauthorblockN{Bowen He}
\IEEEauthorblockA{Zhejiang University\\
bowen\_os@zju.edu.cn}
\\
\IEEEauthorblockN{Yajin Zhou~\IEEEauthorrefmark{1}~\IEEEauthorrefmark{2}}
\IEEEauthorblockA{Zhejiang University\\
yajin\_zhou@zju.edu.cn}
}



\maketitle

\begin{abstract}
In recent years, a more advanced form of phishing has arisen on Ethereum, surpassing early-stage, simple transaction phishing. This new form, which we refer to as \textit{payload-based} transaction phishing (\shortphish{}), manipulates smart contract interactions through the execution of malicious payloads to deceive users. \shortphish{} has rapidly emerged as a significant threat, leading to incidents that caused losses exceeding \$70 million in 2023 reports. Despite its substantial impact, no previous studies have systematically explored \shortphish{}.

In this paper, we present the first comprehensive study of the \shortphish{} on Ethereum. 
Firstly, we conduct a long-term data collection and put considerable effort into establishing the first ground-truth \shortphish{} dataset, consisting of 5,000 phishing transactions. Based on the dataset, we dissect \shortphish{}, categorizing phishing tactics into four primary categories and eleven sub-categories.
Secondly, we propose a rule-based multi-dimensional detection approach to identify \shortphish{}, achieving an F1-score of over 99\% and processing each block in an average of 390 ms.
Finally, we conducted a large-scale detection spanning 300 days and discovered a total of 130,637 phishing transactions on Ethereum, resulting in losses exceeding \$341.9 million. Our in-depth analysis of these phishing transactions yielded valuable and insightful findings.
Scammers consume approximately 13.4 ETH daily, which accounts for 12.5\% of the total Ethereum gas, to propagate address poisoning scams. Additionally, our analysis reveals patterns in the cash-out process employed by phishing scammers, and we find that the top five phishing organizations are responsible for 40.7\% of all losses.

Furthermore, our work has made significant contributions to mitigating real-world threats.
We have reported 1,726 phishing addresses to the community, accounting for 42.7\% of total community contributions during the same period. Additionally, we have sent 2,539 on-chain alert messages, assisting 1,980 victims. 
This research serves as a valuable reference in combating the emerging \shortphish{} and safeguarding users' assets. 
\end{abstract}

\section{Introduction}
The rapid growth of decentralized finance (DeFi) on Ethereum has led to a significant rise in phishing scams. As users actively participate in the DeFi ecosystem, engaging in activities such as purchasing tokens like NFTs and conducting transactions on Ethereum, phishing attempts have adapted to specifically target users' crypto assets.
Unlike traditional phishing scams that focus on privacy or financial information~\cite{bijmans2021catching,kim2022phishing,oest2020sunrise,oest2018inside,oest2020phishtime}, Ethereum phishing are inherently tied to transactions. 
Therefore, we refer to this type of phishing as \textit{transaction phishing} in this paper.

In the early stages, transaction phishing attempts are relatively straightforward, relying on traditional tactics to deceive users. Ethereum transactions are used as a new means of carrying out these scams, rather than being the primary lure for victims. Scammers may initiate transfer transactions through websites to steal victims' crypto assets~\cite{giveawayscam,li2023double}, or entice victims to purchase fake assets~\cite{xia2021trade,ye2023revealing} via websites or crypto wallets. Various mitigation proposals have been suggested to address such threats, such as the detection of phishing websites to limit their spread~\cite{he2023txphishscope,li2023double}, and the prediction of address risk scores based on fund flow relationships~\cite{chen2020phishing,li2022ttagn}.

\begin{table}[!t]
    \centering
    \small
    \caption{Differences between simple transaction phishing and \shortphish{}. \CheckmarkBold{} means the scam has the feature, \XSolidBrush{} means the scam does not.}
    \label{tab:cate_diff}
    \resizebox{.45\textwidth}{!}{%
    \begin{threeparttable}
    \begin{tabular}{clccc}
    \toprule
    \multicolumn{2}{c}{\multirow{2}{*}{\textbf{Phishing category}}}& \multicolumn{3}{c}{\textbf{Feature}}      \\ \cline{3-5} 
    \multicolumn{2}{c}{}  & \multicolumn{1}{c}{\textsc{U}~\tnote{1}} & \multicolumn{1}{c}{\textsc{T}~\tnote{2}} & \textsc{C}~\tnote{3} \\ \hline
    \multicolumn{1}{c}{\multirow{2}{*}{\begin{tabular}[c]{@{}l@{}}Simple transaction \\ phishing\end{tabular}}} & Direct-transfer                                                                 & \multicolumn{1}{c}{\XSolidBrush}                  & \multicolumn{1}{c}{\XSolidBrush}                                                                          & \XSolidBrush                                                            \\ \cline{2-5}
    \multicolumn{1}{c}{} & Fake token purchase & \multicolumn{1}{c}{\XSolidBrush}& \multicolumn{1}{c}{\XSolidBrush}  & \CheckmarkBold   \\ \hline
    \multicolumn{1}{c}{\multirow{4}{*}{\begin{tabular}[c]{@{}l@{}}Payload-based \\ transaction phishing \\ (\shortphish{})\end{tabular}}}      & Ice phish      & \multicolumn{1}{c}{\CheckmarkBold}       & \multicolumn{1}{c}{\CheckmarkBold} & \XSolidBrush    \\ \cline{2-5} 
    \multicolumn{1}{c}{} & NFT order      & \multicolumn{1}{c}{\CheckmarkBold}       & \multicolumn{1}{c}{\CheckmarkBold} & \XSolidBrush    \\ \cline{2-5} 
    \multicolumn{1}{c}{} & Address poison& \multicolumn{1}{c}{\CheckmarkBold}       & \multicolumn{1}{c}{\CheckmarkBold} & \CheckmarkBold   \\ \cline{2-5} 
    \multicolumn{1}{c}{} & Payable function       & \multicolumn{1}{c}{\CheckmarkBold}       & \multicolumn{1}{c}{\CheckmarkBold} & \CheckmarkBold   \\ \hline
    \end{tabular}%
    
    \tnote1[\textsc{U}] \textbf{Web3 unique phishing tactics.} Web3 unique phishing tactics arise from the EVM's design, leveraging smart contracts semantics for phishing rather than simply transferring funds.
    
    \tnote2[\textsc{T}] \textbf{Malicious transaction payload.} Malicious transaction payload refers to transactions with malicious input data executing specific phishing-related smart contracts.
    
    \tnote3[\textsc{C}] \textbf{Malicious contracts deployed by scammers.}
    \end{threeparttable}
    }
    \vspace{-1em}
    \end{table}

However, with the continuous evolution of phishing tactics, more sophisticated scams are emerging that exploit complex on-chain semantics. 
These sophisticated scams involve scammers crafting transactions or messages~\footnote{The signed messages are initially dispatched to the scammer, who subsequently broadcasts them to the blockchain.} that \textit{manipulate smart contract interactions through the execution of malicious payloads} to deceive users. 
These payloads can either be embedded within the malicious smart contracts deployed by the scammers or executed by benign smart contracts used as the executor. 
In this paper, we refer to these scams as \textit{Payload-based Transaction Phishing} (\shortphish{}).
Table~\ref{tab:cate_diff} provides a summary of the differences between the aforementioned simple transaction phishing and \shortphish{}, with a further categorization of \shortphish{} discussed in Section~\ref{subsec:categorization}. 

\begin{figure}[!t]
\centering
	\includegraphics[width=.45\textwidth]{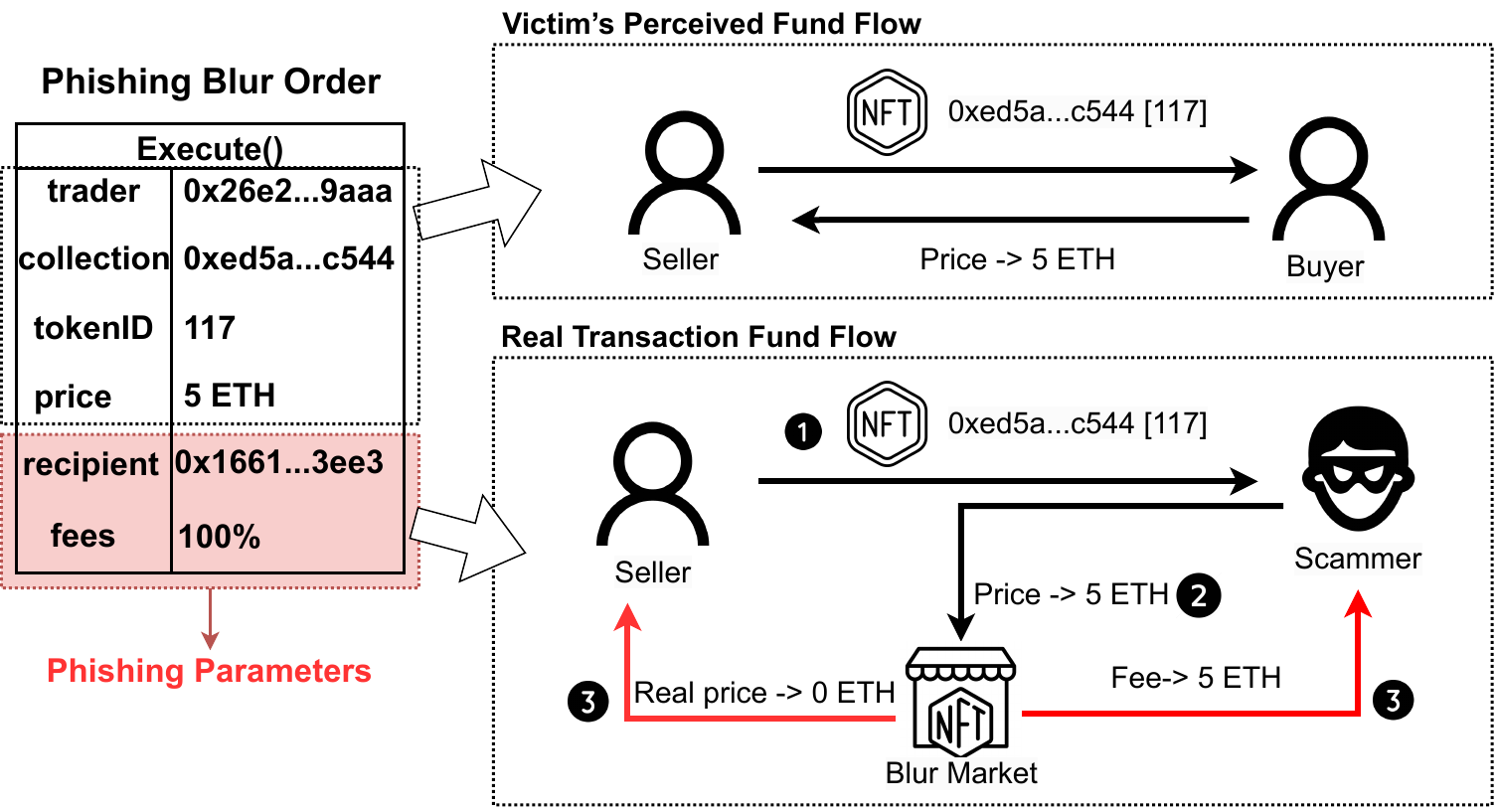}
    \caption{
    A \shortphish{} example that leverages Blur order transaction semantics. From the perspective of the NFT seller, it seems as if a regular buyer is purchasing the NFT for 5 ETH.
    However, the scammer cleverly sets the \code{fees} parameter to 100\% and designates himself as the \code{recipient}. 
    In reality, the seller sends the NFT to the scammer~\ding{182}, and the scammer sends 5 ETH to Blur first~\ding{183}. But due to the 100\% fees, Blur redirects the 5 ETH (calculated as \textit{price * fees}) back to the scammer, who is the designated fee recipient, and sends the remaining 0 ETH (calculated as \textit{price * (1-fees)}) to the seller~\ding{184}. As a result, the scammers appropriate the victims’ NFT without making any payment.}
	\label{fig:blur-fee}
    \vspace{-1em}
\end{figure}

Figure~\ref{fig:blur-fee} provides a \shortphish{} example of a malicious payload executed by a benign smart contract. 
The scammer manipulates the semantics of the Blur~\footnote{\url{Blur.io} is one of the top NFT marketplaces on Ethereum.} order transactions to deceive the victim. The intricate transaction semantics make it difficult for users to understand the role of each parameter in the \code{calldata}. Consequently, victims, especially those lacking domain knowledge, may perceive that they are engaging in transactions with a reputable NFT market, while remaining unaware of the concealed malicious behavior within the transaction's parameters (\eg, \code{fees} in Figure~\ref{fig:blur-fee}). This lack of awareness leads victims to place blind trust in the scammer, ultimately allowing the scammer to successfully appropriate the victim's NFT without making any payment. Additionally, the propagation process and tricks are detailed in Section~\ref{subsubsec:market}.

\shortphish{} has been increasingly prevalent in the recent two years. For example, a significant number of \shortphish{} incidents were reported from November 2022 to July 2023~\cite{Drainer, Btcappr,Bigphisher,OpenSeanew,approve1,approve2,permitReport1,permitReport2,dustTransfer,emptyFunction,emptyFunction2,bulktransfer,Claimphish}, resulting in cumulative financial losses exceeding \$70 million. One particular \shortphish{} incident stands out, causing a loss of \$24 million and ranking among the top ten blockchain attack incidents of 2023~\cite{Largest_phish}. Unfortunately, existing countermeasures have not effectively addressed \shortphish{}, as it exploits transaction semantics in carrying out its scams. Consequently, there is an urgent need to propose an effective detection method to combat \shortphish{}.

Unfortunately, despite the significant threat posed by this emerging type of phishing, the understanding of \shortphish{} is limited. Only a few studies, such as a recent work~\cite{ye2023revealing}, have measured a fraction of \shortphish{}. 
The focus of this particular study~\cite{ye2023revealing} is primarily on visual scams that exploit wallet mistakes, without considering comprehensive contract code. However, it is crucial to perform in-depth contract code analysis to detect transactions that employ sophisticated fraudulent techniques. To the best of our knowledge, no systematic study of \shortphish{} has been conducted to date.

\noindent \textbf{This work.}\tab
In this paper, we present the first comprehensive study that dissects \shortphish{} on Ethereum. We first characterize \shortphish{} and then propose an effective detection approach to combat these scams. Furthermore, we conduct a large-scale and long-term detection and measurement, providing valuable insights into this emerging form of phishing. Our research aims to contribute to the community's understanding and mitigation of such threats.

Specifically, we first conduct extensive data collection and build up the first ground-truth \shortphish{} dataset. Based on this dataset, we propose an in-depth analysis of the processes and tactics used in phishing scams (see Section~\ref{sec:anatomy}). This involves classifying the current \shortphish{} tactics into four main categories and eleven sub-categories.
Drawing from the insights gained through this analysis, we then identify key features of \shortphish{} and propose a rule-based multi-dimensional detection approach accordingly. The effectiveness of this approach in identifying potential \shortphish{} transactions is demonstrated through a thorough evaluation, achieving over 99\% F1-score and processing each block in 390 ms on average. (see Section~\ref{sec:detection}). 
Lastly, we conduct a large-scale detection and perform an extensive analysis of \shortphish{} from three perspectives (see Section~\ref{sec:ana}):
\begin{itemize}[leftmargin = *]
\item[$\circ$]
\textit{The transactions}: We delve into phishing transactions, examining the extent of funds lost and providing detailed insights into the characteristics of each category.
\item[$\circ$]
\textit{The scammers}: We categorize scammer addresses into three types based on their behaviors: \textit{cashiers}, \textit{fund aggregators}, and \textit{depositors}. Additionally, we propose an algorithm based on \textit{cash-out} patterns, which utilizes the relationship between funds and address types to identify and track scammer organizations.
\item[$\circ$]
\textit{The victims}: We scrutinize the profiles of victims, including their address behavior features and the remedial actions they took after falling victim to phishing scams.
\end{itemize}

\noindent \textbf{Our findings.}\tab
In this study, we provide valuable insights into the characteristics of \shortphish{}. Our analysis of \shortphish{} transactions reveals the increasing prevalence of this type of phishing. From December 31, 2022, to October 27, 2023, the frequency of \shortphish{} escalated, resulting in significant economic damage exceeding \$341.9 million across 130,637 transactions. Notably, approximately 4.97\% of \code{approve} transactions and 46.22\% of \code{permit} transactions are identified as phishing transactions.
Our investigations suggest that the NFT markets prove ineffective in preventing the sale of stolen NFTs, with the majority of valuable NFTs being cashed out through platforms such as Blur (61.78\%) and OpenSea (21.97\%). Remarkably, scammers spent over 13.4 ETH per day in gas fees to send address poison transactions, accounting for 12.5\% of the total Ethereum gas usage. 

Additionally, our observations indicate a high level of organization among scammers in their cash-out process. By leveraging our cash-out pattern-based algorithm, we successfully identify the current phishing organizations. Interestingly, the top five phishing organizations are responsible for 40.7\% of the total losses. Regarding the victims, our findings reveal that nearly half of them (40.38\%) do not take remedial measures after incurring losses.

\noindent \textbf{Contributions.}\tab
Our study makes the following contributions:
\begin{itemize}[leftmargin=*]
\item \textbf{Anatomy of \shortphish{}.} Through extensive data collection and long-term on-chain monitoring, we systematically analyze the \shortphish{} process and categorize its tactics (Section~\ref{sec:anatomy}).
\item \textbf{First \shortphish{} open-source dataset.} 
We build the first ground-truth \shortphish{} dataset, which encompasses a comprehensive collection of \textbf{5,000} phishing transactions alongside \textbf{13,557} legitimate transactions. We will release it to the community~\footnote{\url{https://github.com/HypoopyH/PTXPhish}}.
\item \textbf{\shortphish{} transaction detection approach.} 
We propose a rule-based multi-dimensional detection approach that effectively and efficiently identifies phishing transactions, achieving an F1-score of over \textbf{99\%} on both the ground-truth dataset and real-world Ethereum transactions from May 1, 2023, to Jun 1, 2023. The average processing time per block is only \textbf{390 ms}.
\item \textbf{In-depth analysis of \shortphish{}.} We conduct a large-scale detection and perform an in-depth analysis of \shortphish{} to provide insightful findings from three perspectives: \shortphish{}  transactions (Section~\ref{subsec:txanalysis}), \shortphish{} scammers (Section~\ref{subsec:scam}), and \shortphish{} victims (Section~\ref{subsec:vic}).
\item \textbf{Contribution to mitigating real-world threats.}
We help mitigate this emerging threat. Specifically, we have reported 1,726 phishing addresses to the community, accounting for \textbf{42.7\%} of the total community contributions in the same period. Moreover, we have sent \textbf{2,539} on-chain alert messages, assisting \textbf{1,980} victims. The community has acknowledged and recognized our efforts to combat phishing attempts and protect individuals from these threats.

\end{itemize}

\input{section/00Background}

\input{section/01Anatomy}

\input{section/02Detection}
\input{section/03transaction_ana}

\input{section/04attacker_ana}
\input{section/05reality_contribution}
\input{section/06Discussion}
\input{section/07Relatedwork}

\input{section/08Conclusion}

\bibliographystyle{IEEEtran}
\bibliography{sample-base}

\newpage

\appendix

\section{Appendix}
The appendix contains charts and figures mentioned in the main text but not displayed due to space constraints.

\subsection{Important  ERC-20/ERC-721 interface}

\begin{figure}[h]
\centering
\begin{lstlisting}[language=Solidity]
// ERC-20
approve(address _spender, uint256 _value)
transferFrom(address _from, address _to, uint256 _value) 
// ERC-721
approve(address _spender,uint256 _tokenId)
setApprovalForAll(address _operator, bool _approved) 
transferFrom(address _from, address _to, uint256 tokenId) 
\end{lstlisting}
    \caption{Important  ERC-20/ERC-721 interface.}
    \label{fig:ercMethod}
\end{figure}

\subsection{Decision on expanding the number of phishing transactions.}
\label{subsec:randomexample}
Due to the extensive transactions associated with the addresses, manually verifying all historical transactions is impractical. Consequently, we employed a sampling method to obtain historical data. This approach involves a trade-off: a larger sample size significantly increases manual effort, while a smaller sample may result in insufficient coverage. 

We analyzed the number of transactions per address and determined the median count to be 43.5. To balance adequate coverage with manageable effort, we chose a threshold of 50 transactions. Detailed information about the addresses is available at: ~\url{https://github.com/HypoopyH/PTXPhish}.

\subsection{Detailed ground-truth dataset of \shortphish{}}

\begin{table*}[h]
    \centering
    \caption{Detailed ground-truth dataset of \shortphish{}.}
    \resizebox{.9\textwidth}{!}{
    \begin{threeparttable}
    \begin{tabular}{|lll|l|l|l|l|}
    \hline
    \multicolumn{3}{|c|}{Category}    & \multicolumn{1}{c|}{Target Assests} & Spread Method    & Our Findings~\tnote{1} & Dataset Num \\ \hline
    \multicolumn{1}{|l|}{\multirow{6}{*}{\begin{tabular}[c]{@{}l@{}}Exploiting legitimate \\ contract\end{tabular}}} & \multicolumn{1}{l|}{\multirow{3}{*}{Ice phishing}}& Approve & ERC20 token & \multirow{6}{*}{Website}     &   -  & 1247 \\ \cline{3-4} \cline{6-7} 
    \multicolumn{1}{|l|}{} & \multicolumn{1}{l|}{}     & Permit  & ERC20 token &&  -  & 814   \\ \cline{3-4} \cline{6-7} 
    \multicolumn{1}{|l|}{} & \multicolumn{1}{l|}{}     & SetApproveForAll    & NFT   &&   -  & 508  \\ \cline{2-4} \cline{6-7} 
    \multicolumn{1}{|l|}{} & \multicolumn{1}{l|}{\multirow{3}{*}{NFT order}}   & Bulk transfer & NFT   &&   -  & 37 \\ \cline{3-4} \cline{6-7} 
    \multicolumn{1}{|l|}{} & \multicolumn{1}{l|}{}     & Proxy upgrade & NFT   && \CheckmarkBold~\cite{proxyupdate}  & 108\\ \cline{3-4} \cline{6-7} 
    \multicolumn{1}{|l|}{} & \multicolumn{1}{l|}{}     & Free buy order& NFT \& ERC20 token   && \CheckmarkBold~\cite{blurzero}~\tnote{2}  & 464 \\ \hline
    \multicolumn{1}{|l|}{\multirow{5}{*}{\begin{tabular}[c]{@{}l@{}}Deploying phishing\\ contract\end{tabular}}}     & \multicolumn{1}{l|}{\multirow{3}{*}{Address poisoning}} & Zero value transfer & ERC20 token & \multirow{5}{*}{Transaction} &  -    & 104 \\ \cline{3-4} \cline{6-7} 
    \multicolumn{1}{|l|}{} & \multicolumn{1}{l|}{}     & Fake token transfer & ERC20 token &&  -   & 100 \\ \cline{3-4} \cline{6-7} 
    \multicolumn{1}{|l|}{} & \multicolumn{1}{l|}{}     & Dust value transfer & ERC20 token && \CheckmarkBold~\cite{dustTransfer}  & 22 \\ \cline{2-4} \cline{6-7} 
    \multicolumn{1}{|l|}{} & \multicolumn{1}{l|}{\multirow{2}{*}{Payable Function}}  & Airdrop function    & ETH   && -  & 788    \\ \cline{3-4} \cline{6-7} 
    \multicolumn{1}{|l|}{} & \multicolumn{1}{l|}{}     & Wallet function     & ETH   &&  -   & 808 \\ \hline
    \multicolumn{3}{|c|}{Benign Transaction}   & - & - & - & 13557 \\ \hline
    \multicolumn{3}{|c|}{\textbf{Total}}   & - & - & - & 18555 \\ \hline
    \end{tabular}
    \begin{tablenotes}
    \item[1] We were the first to discover and report the new fishing tricks.
    \item[2] We were the first to discover the free buy order scam targeting the Blur.io market.
    \end{tablenotes}
    \end{threeparttable}
    }
    \label{tab:anatomy_phish}
    \end{table*}

We have established the first ground-truth \shortphish{} dataset, consisting of 5,000 phishing transactions. The dataset is categorized into various phishing categories, including 2,569 \textit{ice phishing} transactions, 609 \textit{NFT order} transactions, 226 \textit{address poisoning} transactions, and 1,596 \textit{payable function} transactions. 
Detailed information can be found in Table~\ref{tab:anatomy_phish}.

\subsection{Efficiency evaluation of detection approach.}
\begin{figure}[h]
\centering
\includegraphics[width=.45\textwidth]{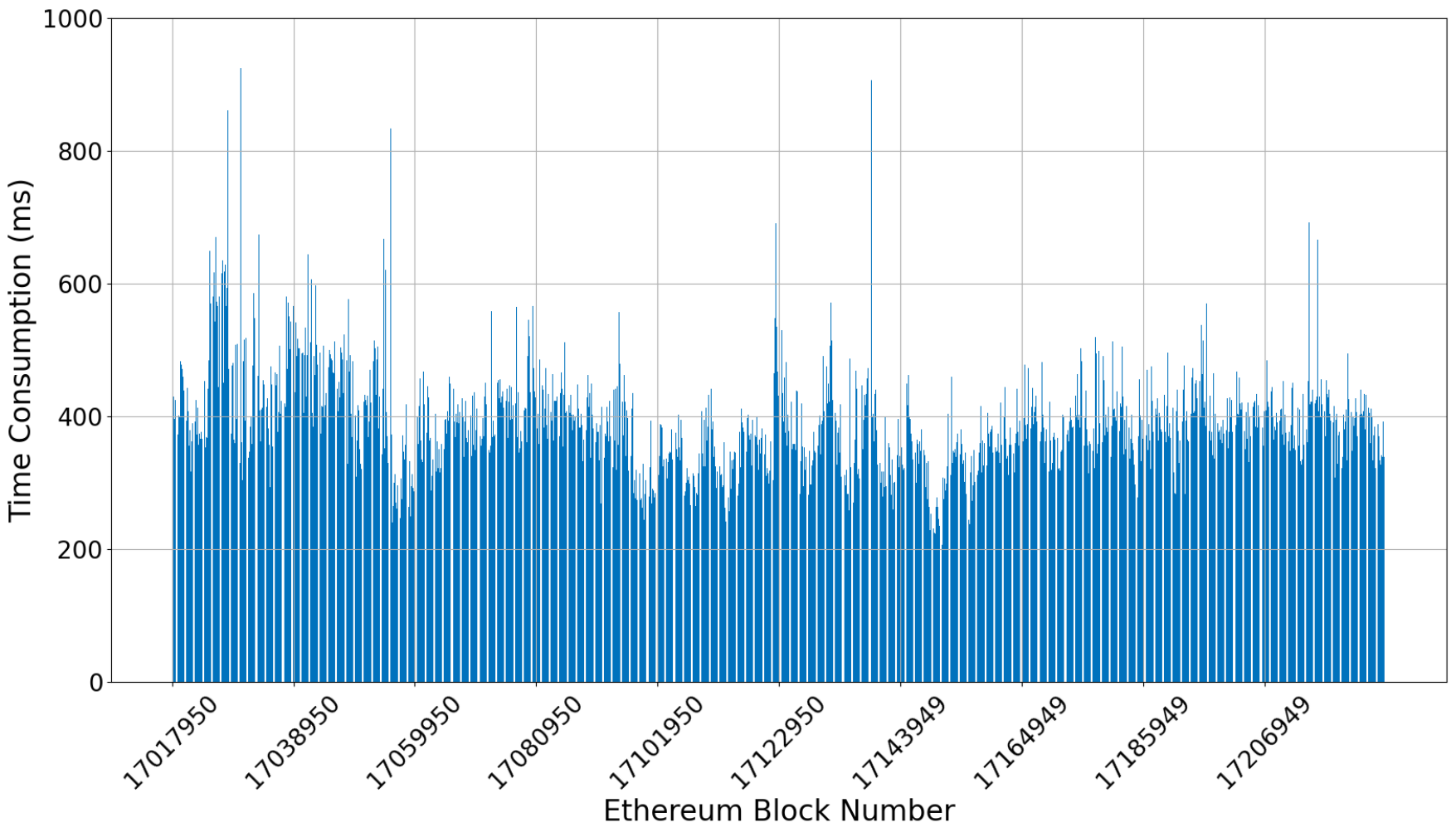}
\caption{The efficiency evaluation of detection methods. Due to the extended duration, the time value is the average time consumption calculated for every ten blocks as a group.}
\label{fig:timeconsumption}
\end{figure}
The figure~\ref{fig:timeconsumption} shows our detection approach time consumption, which is mentioned in Section~\ref{subsec:evaluation}. The average block production time in Ethereum is 12s (12,000 ms). Our approach is highly efficient, with an average time consumption of only 390 ms per block, a median time consumption of 362 ms per block, and a max time of 3,553 ms.

\subsection{Popular signatures of payable function phishing scams}

\begin{table}[h]
    \centering
    \caption{Popular signatures of payable function phishing scams.}
    \resizebox{0.45\textwidth}{!}{
    \begin{threeparttable}
    \begin{tabular}{|ll|l|l|}
    \hline
    \multicolumn{2}{|l|}{Function}& Function Signature & Loss (\$) \\ \hline
    \multicolumn{1}{|l|}{\multirow{3}{*}{Wallet}} & SecurityUpdate & \begin{tabular}[c]{@{}l@{}}0x5fba79f5\\ 0xaf347b61 \\ \end{tabular}&  3,275,537 \\ \cline{2-4} 
    \multicolumn{1}{|l|}{}& ConnectWallet  & 0x62929a1e& 166,302 \\ \cline{2-4} 
    \multicolumn{1}{|l|}{}& NetworkMerge  & 0x9c9316c5& 913,114 \\ \cline{2-4} 
    \multicolumn{1}{|l|}{}& pay& 0x1b9265b8& 419,270 \\ \hline
    \multicolumn{1}{|l|}{\multirow{4}{*}{Airdrop}} & claim/Claim & \begin{tabular}[c]{@{}l@{}}0x4e71d92d\\ 0x3158952e \\0xaad3ec96\\0x0c7ef932 \\ \end{tabular} & 9,717,171  \\ \cline{2-4} 
    \multicolumn{1}{|l|}{}& claimReward & \begin{tabular}[c]{@{}l@{}}0xb88a802f\\ 0x79372f9a \\0xaf7ec6cb \\ 0x63e32091 \\ \end{tabular}& 507,850  \\ \cline{2-4} 
    \multicolumn{1}{|l|}{}& claimRewards& 0xef5cfb8c& 3,456,764 \\ \cline{2-4} 
    \multicolumn{1}{|l|}{}& receiveETH  & 0x4185f8eb& 71,492 \\ \hline
    \multicolumn{2}{|l|}{Total}& -& 18,527,500 \\ \hline
    \end{tabular}
    
    \end{threeparttable}
    }
    \label{tab:contract}
    \end{table}

Table~\ref{tab:contract} described in Section~\ref{subsec:txanalysis}, details popular signatures of payable function phishing scams. These scams have resulted in total losses resulting exceeding \$18 million. We observe two distinct types based on their functionalities: Airdrop scams account for 74.2\% of the total losses (e.g., \code{Claim}/\code{claim}), while Wallet scams account for 25.8\% (e.g.,
\code{SecurityUpdate}).

\subsection{Heatmap of \shortphish{} by date and corresponding losses in the early stage}
\begin{figure*}[!htb]
\centering
	\includegraphics[width=.9\textwidth]{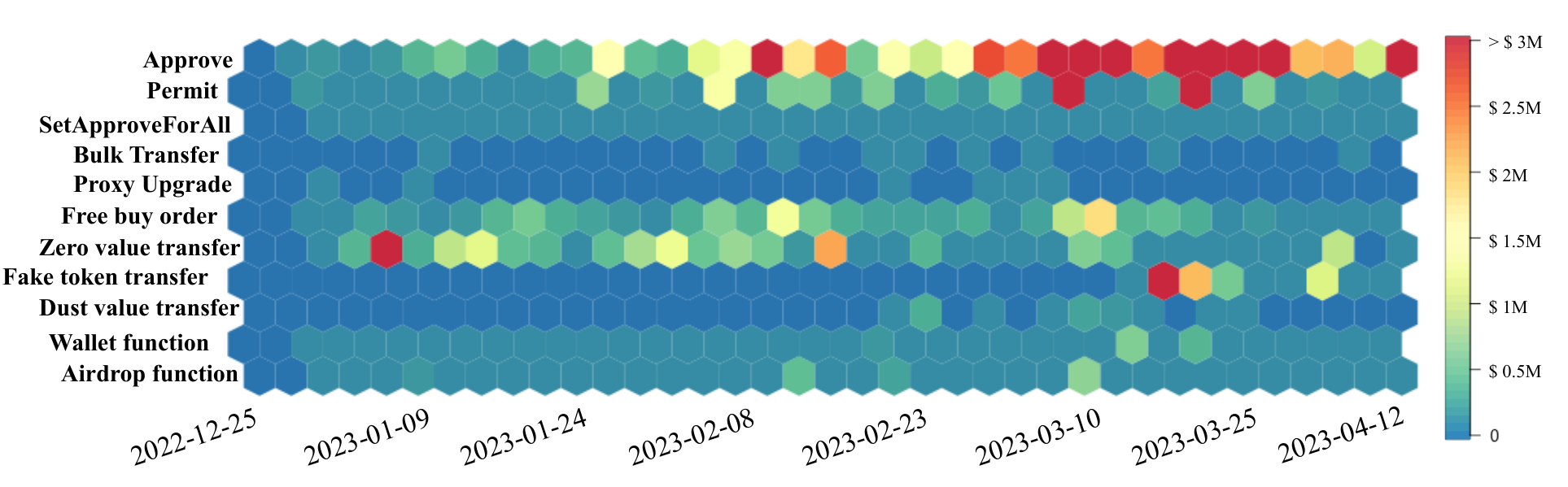}
    \caption{Heatmap of \shortphish{} by date and corresponding losses in the early stage.}
	\label{fig:heatmap}
\end{figure*}
Figure~\ref{fig:heatmap}, described in Section~\ref{subsec:txanalysis} shows the heatmap of \shortphish{} by date and corresponding losses. 
The data indicates that phishing methods are continually evolving and improving. For example, zero value transfer poisoning emerged as an early phishing method on December 25, 2022. However, new variants of address poisoning scams began to appear, with the first successful dust poisoning transaction on March 7, 2023, and the first successful fake token poisoning on March 16, 2023. This timeline highlights the ongoing innovation in phishing techniques.

\subsection{Stolen NFTs cash-out markets}

\begin{table}[h]
    \centering
    \caption{Stolen NFTs cash-out markets.}
    \resizebox{.45\textwidth}{!}{
    \begin{tabular}{llllll}
    \toprule
    \multirow{2}{*}{Seller} & \multicolumn{4}{c}{Markt} & \multirow{2}{*}{Total} \\ \cmidrule{2-5}
       & Blur & Opensea & LooksRare & X2Y2 &  \\ \midrule
    Cashier  & 6,654 & 1,694     & 944& 939    & 10,231     \\ \midrule
    Fund aggregator  & 1,487  & 1,201     & 88 & 158   & 2,934     \\ \midrule
    Total     & 8,141 & 2,895    & 1,032      & 1,097  & 13,165     \\ \bottomrule
    \end{tabular}
    }
    \label{tab:nft_market}
    \end{table}

Table~\ref{tab:nft_market}, described in Section~\ref{subsec:txanalysis}, presents data on stolen NFTs cash-out markets. 
The table reveals that the majority of scammers (62.22\%) directly sell the NFTs to the market using the cashier address, while a smaller portion (17.85\%) transfers NFTs to fund aggregators for selling. 
Among the stolen NFTs, most were sold through Blur (61.78\%), followed by OpenSea (21.97\%), X2Y2 (8.32\%), and LooksRare (7.83\%).

\subsection{Victim behavior profile}
\begin{figure}[h]
\centering
\includegraphics[width=.45\textwidth]{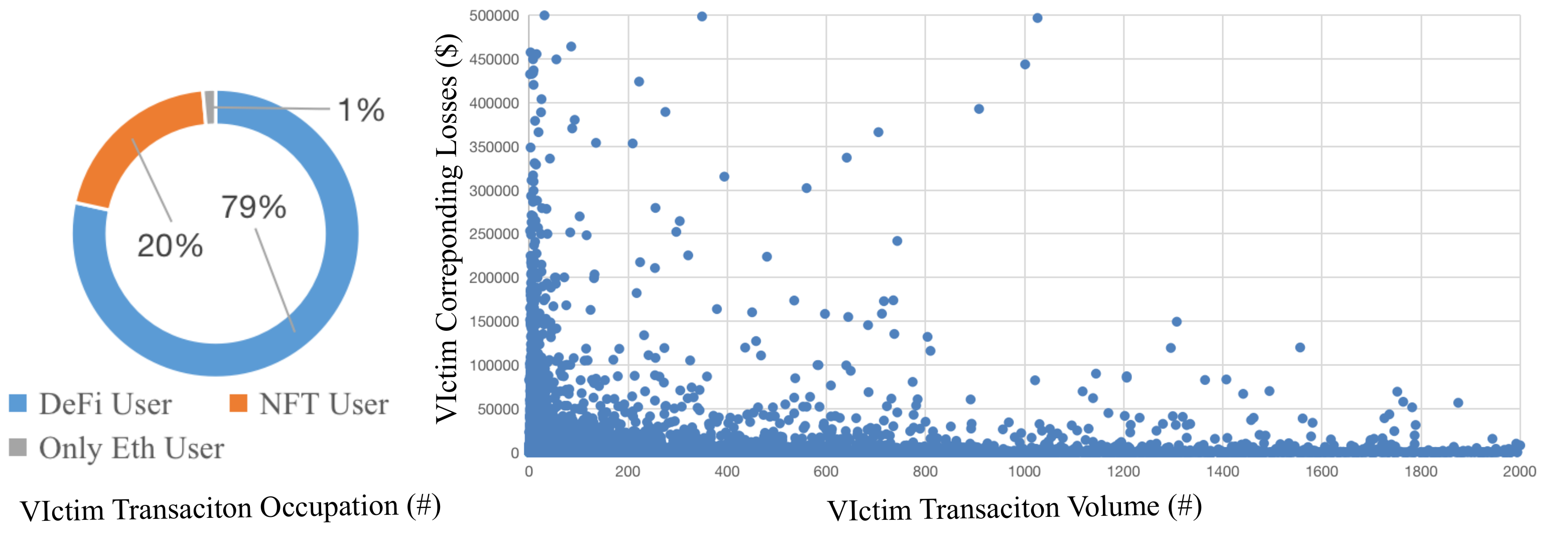}
\caption{The victim behavior profile. Figure 1 is the proportion of victims' transaction types, and Figure 2 is the victims' transaction volume and corresponding losses.}
\label{fig:txloss}
\end{figure}
Figure~\ref{fig:txloss}, mentioned in Section~\ref{subsec:vic}, illustrates the victim behavior profile.
The data shows that the majority of victims have conducted fewer than 1,000 transactions. Notably, in incidents involving large amounts (over \$100k), the victim transactions are predominantly fewer than 50. This suggests that experienced users, who handle higher transaction amounts, are generally more aware of phishing prevention.

\subsection{Scammer organization discovery algorithm process}

\begin{figure}[h]
\centering
\includegraphics[width=.45\textwidth]{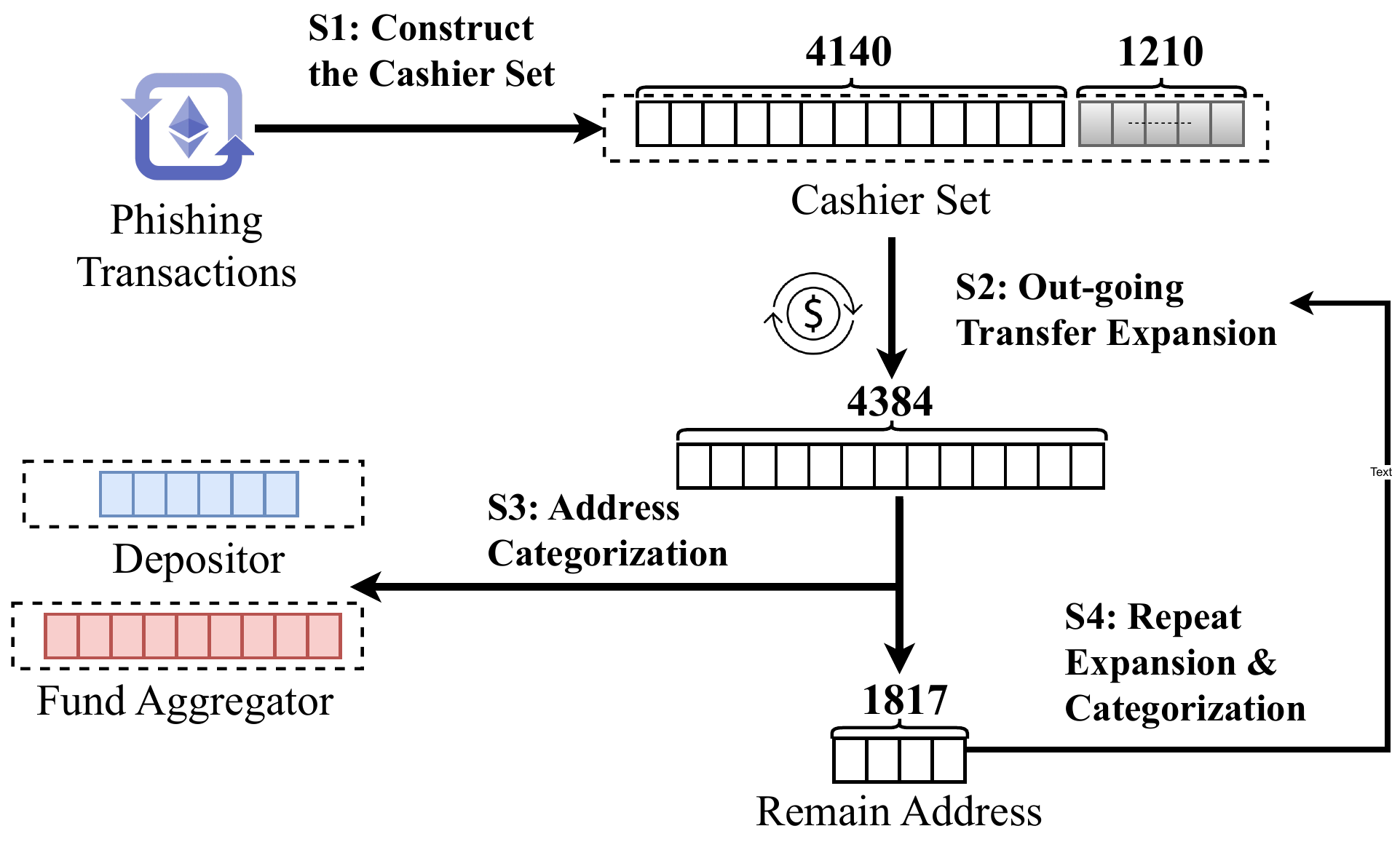}
    \caption{Scammer organization discovery algorithm process.}
	\label{fig:familyturn}
\end{figure}
Figure~\ref{fig:familyturn}, described in Section~\ref{subsec:scam}, outlines the scammer organization discovery algorithm process. 
In step S1, we found 5,350 cashier addresses, of which 1,210 had no outgoing transfers.
In step S2, we identified 4,384 outgoing destination addresses with transfer value exceeding 100\$. 
In step S3, we categorized these addresses into 2,307 destination fund aggregators and 260 depositors based on their behavior. 
In step S4, we repeated the outgoing transfer expansion \& categorization process for the remaining 1,817 addresses.

\subsection{Details of existing anti-phishing tools/platforms}

\begin{table}[h]
    \centering
    \caption{Details of existing anti-phishing tools/platforms. \CheckmarkBold means the tool/platform leverages the feature, \XSolidBrush means the tool/platform does not.}
    \resizebox{.45\textwidth}{!}{

    \begin{threeparttable}
    \begin{tabular}{lcccl}
    \toprule
    \multirow{2}{*}{ \textbf{Category}}& \multirow{2}{*}{\textbf{Tool/Platform}}       & \multirow{2}{*}{\begin{tabular}[c]{@{}l@{}}\textbf{Blacklist}\\ \textbf{Label}\end{tabular}} & \multirow{2}{*}{\begin{tabular}[c]{@{}l@{}}\textbf{Pre-execution} \\  \end{tabular}} & \multirow{2}{*}{\begin{tabular}[c]{@{}l@{}}\textbf{User} \\ \textbf{Number}\end{tabular}}         \\
       &       &      &          &                \\ \midrule
    \multirow{3}{*}{\begin{tabular}[c]{@{}l@{}}Website \\ Blocking\end{tabular}}    & AegisWeb3        & \CheckmarkBold   & \XSolidBrush      & 100,000+              \\ \cline{2-5}
       & MetaShield       & \CheckmarkBold   & \XSolidBrush      & 1,000+         \\ \cline{2-5}
       & ScamSniffer      & \CheckmarkBold   & \CheckmarkBold      & 40,000+         \\ \midrule
    \multirow{2}{*}{\begin{tabular}[c]{@{}l@{}}Transaction \\ Pre-execution\end{tabular}} & Pocket Universe  & \CheckmarkBold  & \CheckmarkBold      & 100,000+               \\ \cline{2-5}
       & Stelo & \CheckmarkBold  & \CheckmarkBold      & 9,000+          \\ \midrule
    \multirow{2}{*}{\begin{tabular}[c]{@{}l@{}}Transaction \\ Blocking\end{tabular}}    & \multirow{2}{*}{Forta Scam Bot} & \multirow{2}{*}{\CheckmarkBold}       & \multirow{2}{*}{\XSolidBrush}& \multirow{2}{*}{-~\tnote1}     \\
       &       &      &          &                \\ \midrule
    \multirow{2}{*}{\begin{tabular}[c]{@{}l@{}}Remedial \\ Tool\end{tabular}} & Revoke.cash  & \CheckmarkBold  & \XSolidBrush      & 60,000+               \\ \cline{2-5}
       & MetaSleuth.io & \CheckmarkBold  & \XSolidBrush      & 3,500+          \\ \bottomrule
    \end{tabular}
    \begin{tablenotes}
    \small
    \item[1] "-" Represents that the number of users cannot be known.
    \end{tablenotes}
    \end{threeparttable}
    }
    \label{tab:detect_tool}
    \end{table}

Table~\ref{tab:detect_tool}, described in Section~\ref{sec:dis}, provides details on existing anti-phishing tools/platforms. 

\subsection{Remedial behavior of ice phishing victims}

\begin{table}[!t]
    \centering
    \caption{Remedial behavior of ice phishing victims.}
    \resizebox{.45\textwidth}{!}{
    \begin{threeparttable}
    \begin{tabular}{lcc}
    \toprule
    \multirow{2}{*}{\textbf{Remedial measure}} & \multirow{2}{*}{\textbf{Number (\#)}} & \multirow{2}{*}{\textbf{Proportion (\%)}} \\
     && \\ \midrule
    Revoke & 1,316 & 26.32\%  \\ \midrule
    Transfer assets & 1,665 & 33.3\%  \\ \midrule
    No remedial measure& 2,019 & 40.38\%  \\ \midrule
    \textbf{Total} & \textbf{5,000} & \textbf{100.0\%}  \\ \bottomrule
    \end{tabular}
    \end{threeparttable}
    }
    \label{tab:revoke}
    \vspace{-1em}
    \end{table}

Table~\ref{tab:revoke}, described in Section~\ref{subsec:vic}, details the remedial behavior of ice phishing victims.
Among the randomly selected 5,000 victims, only 1,316 addresses (26.32\%) chose to revoke the phishing approval, while 1,665 addresses (33.3\%) transferred all funds to other addresses, abandoning the compromised address.
However, a concerning 2,019 addresses (40.38\%) did not take any remedial measures, leaving them vulnerable to further attacks and potential financial loss. 
This suggests that the majority of victims unaware of how to effectively address phishing incidents.






%



\end{document}

%% file: solidity.tex
\usepackage{listings, xcolor}

\definecolor{verylightgray}{rgb}{.97,.97,.97}

\lstdefinelanguage{Solidity}{
	keywords=[1]{anonymous, assembly, assert, balance, break, call, callcode, case, catch, class, constant, continue, constructor, contract, debugger, default, delegatecall, delete, do, else, emit, event, experimental, export, external, false, finally, for, function, gas, if, implements, import, in, indexed, instanceof, interface, internal, is, length, library, log0, log1, log2, log3, log4, memory, modifier, new, payable, pragma, protected, pure, push, require, return, returns, revert, selfdestruct, send, solidity, storage, struct, suicide, super, switch, then, throw, true, try, typeof, using, value, view, while, with, addmod, ecrecover, keccak256, mulmod, ripemd160, sha256, sha3}, 
	keywordstyle=[1]\color{blue}\bfseries,
	keywords=[2]{MAX\_UINT, bool, byte, bytes, bytes1, bytes2, bytes3, bytes4, bytes5, bytes6, bytes7, bytes8, bytes9, bytes10, bytes11, bytes12, bytes13, bytes14, bytes15, bytes16, bytes17, bytes18, bytes19, bytes20, bytes21, bytes22, bytes23, bytes24, bytes25, bytes26, bytes27, bytes28, bytes29, bytes30, bytes31, bytes32, enum, int, int8, int16, int24, int32, int40, int48, int56, int64, int72, int80, int88, int96, int104, int112, int120, int128, int136, int144, int152, int160, int168, int176, int184, int192, int200, int208, int216, int224, int232, int240, int248, int256, mapping, string, uint, uint8, uint16, uint24, uint32, uint40, uint48, uint56, uint64, uint72, uint80, uint88, uint96, uint104, uint112, uint120, uint128, uint136, uint144, uint152, uint160, uint168, uint176, uint184, uint192, uint200, uint208, uint216, uint224, uint232, uint240, uint248, uint256, var, void, ether, finney, szabo, wei, days, hours, minutes, seconds, weeks, years},	
	keywordstyle=[2]\color{teal}\bfseries,
	keywords=[3]{block, blockhash, coinbase, difficulty, gaslimit, number, timestamp, msg, data, gas, sender, sig, value, now, tx, gasprice, origin},	
	keywordstyle=[3]\color{violet}\bfseries,
	identifierstyle=\color{black},
	sensitive=true,
	comment=[l]{//},
	morecomment=[s]{/*}{*/},
	commentstyle=\color{gray}\ttfamily,
	stringstyle=\color{red}\ttfamily,
	morestring=[b]',
	morestring=[b]"
}

%% file: section/00Background.tex
\section{Background}
\label{sec:background}
\subsection{Ethereum Blockchain}
Ethereum is a public blockchain-based distributed computing platform and operating system featuring scripting functionality. The Ethereum blockchain~\cite{vujivcic2018blockchain} is the most prominent framework for smart contracts~\cite{wu2021defiranger}.

\noindent\textbf{Address.}
In Ethereum, the account can be divided into two types: externally owned account (EOA) and contract account (CA). The EOA is created by using the public-private keys and is controlled by the entity in possession of the private key. 
On the other hand, the CA is created by the EOA through contract creation transactions.
The functionality of CA is controlled by its deployed code instead of an entity. What's more, the CA relies on EOA to execute its functions.

\noindent\textbf{Transaction.} During the operation of Ethereum, users can interact with other users and contracts through sending transactions.
A transaction is a signed message to be sent from an EOA to another account, which carries the following information: to (receiver), from (message sender), value (the amount of native token, \ie, ETH in Ethereum), data (the input for a contract call), \etc. 
In particular, when a transaction sets its \textit{to} field to be empty, Ethereum regards it as a transaction that creates a contract with its data field being the bytecode of the contract. 
In the end, transactions will be verified by all chain clients and be written onto the blockchain.

\subsection{Decentralized Finance (DeFi)}
Decentralized Finance (DeFi) is an emerging model for organizing and enabling cryptocurrency-based transactions~\cite{wu2021defiranger}. In Ethereum, DeFi is built on top of multiple smart contracts, giving rise to projects such as lending, trading, and marketing~\cite{zhou2021just,wang2021towards,werner2022sok}.

\noindent \textbf{Token.}
In Ethereum chains, tokens are digital assets. Unlike native cryptocurrency (\ie, ETH in Ethereum), tokens are implemented using specialized smart contracts. There are two main types of tokens: fungible and non-fungible.

\lstset{
 columns=fixed,       
 numbers=left,                                        
 numberstyle=\fontzise\color{gray},                       
 frame=none,                                          
 backgroundcolor=\color[RGB]{245,245,244},            
 keywordstyle=\color[RGB]{40,40,255},                 
 numberstyle=\footnotesize\color{darkgray},           
 commentstyle=\it\color[RGB]{0,96,96},                
 stringstyle=\rmfamily\slshape\color[RGB]{128,0,0},   
 showstringspaces=false,                              
 language=python,                                        
 breaklines = true, 
 basewidth       =   0.5em
}

Fungible tokens, which are homogeneous and interchangeable, mostly conform to the same interface standard. These tokens serve as a complement to the native currency, playing the role of a more flexible secondary currency within the DeFi ecosystem.
In contrast, non-fungible tokens (NFT) conform to a different type of interface, such as ERC-721/1155 in ETH. These tokens are identified with unique \code{\_tokenID}, representing a digital asset such as ENS domains or pictures.

ERC-20/721 are currently the most widely used standards for token implementation on Ethereum.
The standard interface defines a set of API methods that a token contract needs to implement.  
Some important API methods relevant to our study are listed in Figure~\ref{fig:ercMethod} (in the appendix).
The \code{approve} method approves the \code{\_spender} as the operator of the token (msg.receiver) with \code{\_value} (ERC-20) or \code{\_tokenId} (ERC-721). 
In ERC-721, the \code{setApprovalForAll} method can either add or remove the address \textit{\_operator} from/to the set of the operators authorized by the msg.sender.
The spender can call the \code{transferFrom} method to transfer the token (within the approve \code{\_value} in ERC-20, or the same \code{\_tokenId} in ERC-721) from the current owner’s \code{\_from} address to the \code{\_to} address.

\noindent\textbf{NFT Marketplaces.}
NFT marketplaces are decentralized application (dApp) platforms where NFTs are traded. Typically, there are two main components of an NFT marketplace: a user-facing web interface and a collection of smart contracts that interact with the blockchain. Users interact with the web app, which in turn sends transactions to the smart contracts. To facilitate these transactions, these marketplaces have implemented many methods to help users place orders, make purchases, and transfer NFTs in batches.

%% file: section/01Anatomy.tex
\section{Anatomy of \shortphish{}}
\label{sec:anatomy}
In this section, we first describe our data collection process of the \shortphish{} dataset. 
We then analyze phishing tactics and categorize current phishing scams. Finally, we evaluate the coverage and effectiveness of our anatomy.

\subsection{Data Collection of \shortphish{}}
\label{subsec:informaton_collect}

Currently, there is no centralized source of information dedicated to \shortphish{}, and public information sources are diverse. 
To address this gap, we have created the first ground-truth phishing transaction dataset. 
Specifically, our dataset was established through the following steps:
\begin{itemize} [leftmargin=*]
\item \textbf{Collecting public reports.} 
We gathered public reports from two sources, \ie, the phishing complaints made by victims on social media and the phishing blogs reported by the security community~\cite{Forta_blog,BlockSec_blog,Peckshiled_blog,Metasleuth_blog,ScamSniffer_blog}. 
By querying keywords related to \textit{phishing}, \textit{scam}, and \textit{drainer}, we identified relevant websites. Our information collection lasted for three months and resulted in 101 public phishing complaints and reports.

\item \textbf{Reviewing public reports.}
Due to the diverse sources of phishing reports, these phishing reports are in different formats and lack authoritative verification. 
To ensure the accuracy of the dataset, a manual review was conducted by two security experts. They analyzed the transaction data, logs, tokens transferred, and transaction call traces. A consensus was reached by the two experts to label a transaction as phishing.
During the review process, we recorded scammers' and victims' addresses, transaction parameters, and transaction hashes to standardize the data format.

\item \textbf{Expanding from the historical phishing data.} 
To increase the number of phishing transactions, we reviewed the transaction history of the scammers' addresses collected from the public reports. Random historical transactions were selected from each scammer's address, with an additional 50 transactions chosen for each phishing address~\footnote{Our investigation suggests that a threshold of 50 is typically enough to cover the majority of phishing techniques, see Appendix~\ref{subsec:randomexample} for details.}. The extended transactions underwent manual review as in the previous step. 
Notably, through our data extension and manual reviews, we have found some hidden phishing scams and provided several first-of-its-kind reports of new scams (\ie, Blur free buy order, dust value poison).
\end{itemize}

By doing so, we have established the first ground-truth \shortphish{} dataset, which consists of 5,000 phishing transactions. The \shortphish{} dataset is further categorized into different phishing categories (see Section~\ref{subsec:categorization}), including 2,569 \textit{ice phishing} transactions, 609 \textit{NFT order} transactions, 226 \textit{address poisoning} transactions, and 1,596 \textit{payable function} transactions. 
The detailed information can be found in Table~\ref{tab:anatomy_phish} in the appendix due to the page limit.

Furthermore, we built a benign dataset for comparison by collecting transactions from two distinct sources:
\begin{itemize} 
\item 
Top 50 Debank~\footnote{A well-known website for tracking Web3 portfolio~\cite{DEBANK}.} Key Opinion Leaders (KOL). 
These influential users significantly impact the investment community and have a large following, which bolsters the credibility of their transactions.
\item
Top 10 DeFi Protocol Developers~\footnote{Based on DefiLlama~\cite{DefiLlma}, a top site for DeFi project rankings.}.
These high-level developers are prominent in the DeFi space, and their widely used contracts underscore the legitimacy of their transactions.
\end{itemize}
To ensure comprehensive representation and maintain a balanced sample size, we randomly selected 200 transactions for each user~\footnote{Addresses with fewer than 200 transactions were included in full.}. 
In total, we gathered 13,557 benign transactions.

\subsection{Categorization of \shortphish{}}
\label{subsec:categorization}
Based on the ground-truth dataset, our analysis reveals that scammers employ two distinct strategies: \textit{(i)} Abusing legitimate contracts; and \textit{(ii)} Exploiting phishing contracts, as depicted in Figure~\ref{fig:type-1-phish}. 
In the following, we delve into the details of these strategies, including their progress and specific tactics. 

\subsubsection{Abusing legitimate contract}
\begin{figure}[!t]
\centering
\includegraphics[width=.45\textwidth]{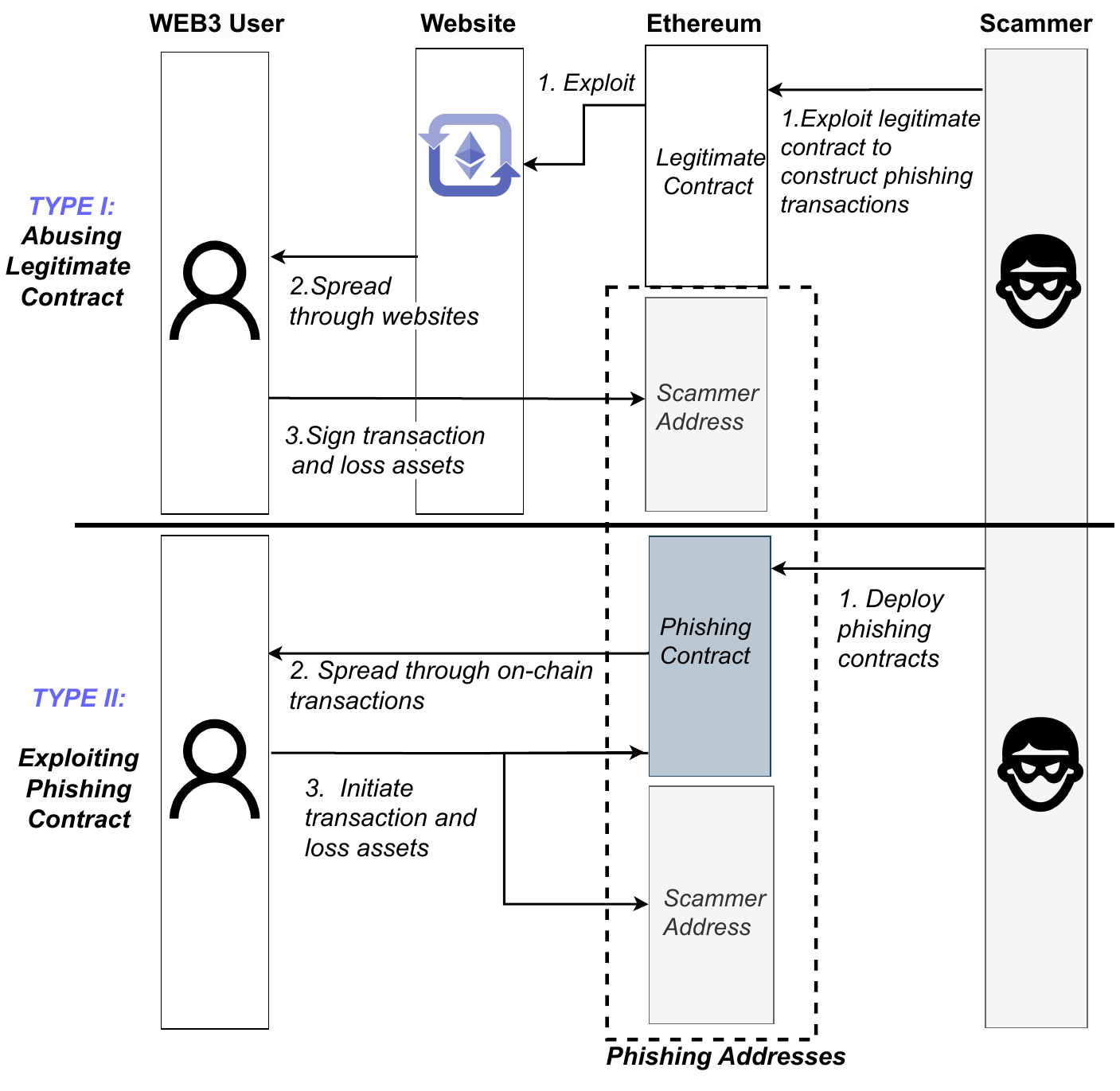}
    \caption{Anatomy of \shortphish{}. According to the strategies, \shortphish{} is divided into two types: (i) Abusing legitimate contracts. (ii) Exploiting phishing contracts. }
	\label{fig:type-1-phish}
    \vspace{-1em}
\end{figure}

As depicted in Figure~\ref{fig:type-1-phish}, abusing legitimate smart contracts involves three steps.
We provide a thorough description of them in the following:
\begin{itemize}[leftmargin=*]
    \item[$\circ$] \textit{Step I: Scammer abuses legitimate contracts to construct phishing transactions.} At first, the scammer analyzes well-known DeFi projects' contracts (\eg, ERC20 token contracts, NFT market contracts, and Uniswap contracts) and their functions.
    Subsequently, based on some functions of these contracts, the scammer constructs a set of transactions with malicious semantics. Although the interaction targets of these transactions are legitimate contracts, their actual behavior will cause phishing scams.
    
    \item[$\circ$] \textit{Step II: Scammer spreads phishing transactions through websites.}  Generally, the scammer conceals phishing transactions within fraudulent websites and promotes them on social media platforms such as Twitter, Telegram, Instagram, or Discord.
    When visiting fake websites, victims would connect their wallet and be asked to sign a transaction~\footnote{Since the user's address is different, the phishing website will adjust the phishing transaction request according to the user's address.}.
    Unfortunately, victims only understand they are interacting with authorized contracts but are unaware of the real output of the phishing transactions, leading them to place blind trust in the phishing transactions.

    \item[$\circ$] \textit{Step III: Victims sign phishing transactions and lose assets.} 
    Once the victim signs and submits the transaction to the Ethereum client, the legitimate contract will execute it, transferring/authorizing the victim's assets to the scammer.
\end{itemize}

The essence of abusing existing legitimate contracts involves deceiving victims by making them believe the transactions conducted with authoritative contracts are legitimate.
Therefore, based on the types and methods of the exploited legitimate contracts, we can divide them into two categories: \textit{ice phishing} and \textit{market order} scams.

\noindent\textbf{Scam Category I: Ice phishing scam.}
\label{subsubsec:ice}
The ice phishing scam exploits the \code{approve} function in token contract (see Section~\ref{sec:background}). 
Token owners can call the \code{approve} function to give an address the right to control a certain amount of their tokens. 

However, the interface does not impose any limitations on the spender.
Specifically, \textit{(i)} the spender can be any address, no matter whether it is a Contract Account (CA) or an Externally Owned Account (EOA); \textit{(ii)} the spender has the ability to transfer approved amount of tokens to any other addresses.  
In other words, if the spender is an EOA address, it can arbitrarily transfer the owner's assets to any address without the owner's consent. 
There are three specific sub-categories:

\begin{itemize}[leftmargin=*]
    \item[$\diamond$] \textit{I-A: Approve.}
    Targeting victims' ERC-20 tokens, the scammer constructs phishing transactions with \code{approve} (ERC-20 standard interface) and \code{increaseAllowance} (optional ERC-20 interface) to lure victims to sign. 
    
    \item[$\diamond$] \textit{I-B: Permit.}
    The \code{permit} function performs the same role as the \code{approve} function but allows for off-chain signing. 
    Exploiting this feature, the scammer creates off-chain ERC20 \code{permit} messages and lures victims into signing them.
    The scammer then submits the \code{permit} transaction to Ethereum.
    \item[$\diamond$] \textit{I-C: SetApproveForAll.}
    Turning to NFTs, the scammer exploits the setApproveForAll function of NFT collections, which can approve an entire NFT collection to an address within a single transaction.
\end{itemize}

\noindent\textbf{Scam Category II: NFT order scam.}
\label{subsubsec:market}
NFT order scams specifically target popular NFTs owned by victims. Since the majority of users manage and trade their NFTs through dedicated NFT markets such as OpenSea~\cite{Opensea} and Blur~\cite{blur}, scammers abuse the existing NFT market contracts to construct deceptive transactions.

Due to the lack of unified interfaces, NFT markets have implemented their own market order contracts. These contracts are highly complex, making it challenging for users to comprehend the corresponding transactions. Even wallets are only able to display raw data without providing clear explanations. Consequently, we have observed three commonly employed tactics in these scams.
\begin{itemize}[leftmargin=*]
    \item[$\diamond$] \textit{II-A: Bulk transfer.} 
     Aiming to simplify the process of transferring multiple NFTs to a designated recipient address, OpenSea introduced a convenient function called the \code{bulkTransfer}.
     Regrettably, scammers exploit this function by surreptitiously replacing the intended \code{recipient} address with their own, thereby diverting the NFTs to their control.

    \item[$\diamond$] \textit{II-B: Proxy upgrade.}
    In the early stage, OpenSea implemented a proxy contract to streamline the trading process for its users. By default, this proxy contract initially grants operator rights over the user's NFTs. 
    Exploiting this feature, scammers deceive users into signing a \code{proxy upgrade} transaction, which replaces the proxy contract's implementation with a scammer-controlled contract. As a result, the scammers gain ownership of the proxy contract~\cite{bodell2023proxy}, thereby enabling them to steal the user's NFTs through the manipulated proxy contract.
    
    \item[$\diamond$] \textit{II-C: Free buy order.}
    In contrast to traditional centralized markets, NFT markets utilize a combination of front-end web pages and smart contracts~\cite{das2022understanding}. Specifically, users use the front-end interface to sign an off-chain message that describes their order details, including the floor price and the trade time window. Upon matching the order, the market automatically completes the remaining details, such as the recipient and the final price.

    Exploiting this design, scammers construct transactions with malicious parameters that ultimately result in the loss of the NFT owner. As illustrated in Figure~\ref{fig:blur-fee}, the occurrence of a free buy order is caused by a malicious 100\% \code{fees} parameter intentionally set by the scammer.
\end{itemize}

\begin{figure}[!t]
    \centering
\begin{lstlisting}[language=Solidity]
// transferFrom 0 token => _value = 0
function transferFrom(address _from, address _to, uint _value){
// default _allowance is 0 => _allowance = 0
var _allowance = allowed[_from][msg.sender];
// _allowance - _value = 0 => pass check
if (_allowance < MAX_UINT) {
    allowed[_from][msg.sender] = _allowance.sub(_value);} 
    ...
// _from transfer 0 token to _to
Transfer(_from, _to, _value);}

\end{lstlisting}
    \caption{Simplified \code{transferFrom} function of the USDT token. Any zero value transfer between two addresses is permitted.}
    \label{fig:usdtcode}
    \vspace{-1em}
\end{figure}

\subsubsection{Exploiting phishing contracts}
As depicted in Figure~\ref{fig:type-1-phish}, the process of exploiting malicious contracts deployed by scammers involves three steps.
We provide a detailed description of each step below:
\begin{itemize}[leftmargin=*]
    \item[$\circ$] \textit{Step I: Scammer deploys phishing contracts.} In this kind of scam, scammers begin by deploying one or a group of contracts with different functionalities. Common malicious contracts include fake token, broadcoast, and trap contracts.
    
    \item[$\circ$] \textit{Step II: Phishing contracts spread fake information to victims through transactions.}  
     In contrast to spreading scams through websites, scammers employ broadcast contracts to spread fake information to users through on-chain transactions. These transactions are specially designed to contain false information that can contaminate users' wallets. For example, they can poison users' transaction records or airdrop tokens with fake information.
    
    \item[$\circ$] \textit{Step III: Victims believe the fake information, initiate transactions and lose assets.} The victims believe the information appearing in their wallets and initiate transactions to phishing addresses.  Unfortunately, these user-initiated transactions lead to the loss of their assets.

\end{itemize}

According to the different contracts the scammers deployed, we can divide them into two categories: \textit{address poisoning},
and \textit{payable function} scam.

\noindent\textbf{Scam Category III: Address poisoning scam.}
\label{subsubsec:address}
The address poisoning scam is a distinct type of scam within the blockchain ecosystem. Its primary objective is to create fake transactions between fake addresses and user addresses actively. By doing so, the scammer effectively contaminates the user's transaction records with these fake addresses~\footnote{The full length of an address is 20, so the GUI of wallets commonly omits a part of the address, causing similar addresses to display the same.}.

\begin{figure}[!t]
    \centering
\begin{lstlisting}[language=Solidity]
contract SecurityUpdates {
address private owner;
constructor() { owner = msg.sender }
function withdraw() public payable {
 require(msg.sender==owner, "Bro? Are you idiot?")
 payable (msg.sender).transfer(address(this).balance)
}
function SecurityUpdate() public payable {
 if (msg.value > 0) payable(owner).transfer(address(this).balance)
}}

\end{lstlisting}
    \caption{An example of a malicious \code{SecurityUpdate} function. This fraudulent implementation has the \code{payable} modifier to receive the victim's native tokens. When victims attempt to withdraw their funds, the scammer will mock them.}
    \label{fig:securityUpdate}
    \vspace{-1em}
\end{figure}

For ease of understanding, we show a famous address poison scam example~\cite{Largest_phish_binance}. Initially, Binance sent 10 million USDT to a legitimate deposit address (\textit{0xa7B4BAC8f0f9692e56750aEFB5f6cB5516E90570}). 
After monitoring this transfer, the scammer creates a counterfeit address (\textit{0xa7Bf48749D2E4aA29e3209879956b9bAa9E90570}) that has the same GUI (\code{0xa7B4...0570}) in various wallets. 
And then, the scammer \code{TransferFrom} 10 million fake USDT from Binance to the fake address. This action leaves a record of the fake address in Binance's transfer history.
In a crucial misstep, Binance mistakenly believed the fake address in transfer history and transferred another 20 million USDT to the fake address in transaction~\footnote{Transaction hash: 0x08255ca0e42a872559437141fa46980e66d907f7668922\\467d67515b1ebb4b7f}.  This mistake behavior results in the loss of the funds.

The scam exploits the victims who mistakenly believe that all the history records are initiated by themselves. Specifically, there are three sub-categories, as follows:

\begin{itemize}[leftmargin=*]
    \item[$\diamond$] \textit{III-A: Zero value transfer.} The interface of ERC-20 tokens specifies that the spender can only transfer tokens within the approved amount. However, by default, the approved amount is set to zero. Exploiting this default behavior, scammers can invoke the \code{transferFrom} function to transfer zero tokens from the victim's address to a fraudulent address, as depicted in Figure 4. Even though no tokens are transferred, this action leaves a transfer record in the victim's transaction history, potentially misleading the victim.
    
    \item[$\diamond$] \textit{III-B: Fake token transfer.}  The scammers deploy fake tokens with the same name/symbol as authoritative tokens. 
    What's more, the scammers remove \code{allowance} check so they can call \code{transferFrom} to transfer any fake tokens from the victim's address to a fake address. By doing so, scammers can leave fake addresses in the victim's transfer history.
    
    \item[$\diamond$] \textit{III-C: Dust value transfer.} The scammer sends a small number of valuable tokens from a fake address to the victim's address. This leaves the fake address in the victim's transfer history. Since they are tiny amounts, they are called dust value transfer.
\end{itemize}

\noindent\textbf{Scam Category IV: Payable function scam.}
\label{subsubsec:mislead}
Due to the absence of an auditing mechanism on the blockchain, the functionality of smart contract functions may not comply with interface protocols but may instead be determined by the smart contract developer.

For better understanding, we show a concrete example in Figure~\ref{fig:securityUpdate}.
The scammer poses as a legitimate project and deceives victims into believing this is a standard interface. 
However, the malicious \code{SecurityUpdate} function accepts the victims' native tokens (via the \code{payable} modifier), while the \code{withdraw} function permits only the ``owner'' of the contract (typically the scammer) to withdraw the tokens. The victim will incur losses upon calling this function with native tokens.

Specifically, there are two major sub-categories, as follows:
\begin{itemize}[leftmargin=*]
    \item[$\diamond$]  \textit{IV-A: Airdrop function.}
     Airdrops are common in DeFi~\cite{Airdrop}. Scammers exploit users' greed and pretend to be an airdrop project. They first airdrop fake tokens to victims and lure victims into calling standard airdrop interfaces, such as the \code{Claim}, \code{ClaimReward}, \code{ClaimRewards}. After victims call the function, they steal victims' native tokens.
    \item[$\diamond$]  \textit{IV-B: Wallet function.}
    Most users use wallets to manage their addresses.
    The scammer pretends to be the user's wallet and sends a message, asking the user to call functions similar to the wallet's functionality and steal their native token. For example, the \code{SecurityUpdate} function pretends to be a wallet update, and the \code{ConnectWallet} function pretends to be a wallet connection.
\end{itemize}

From the analysis of \shortphish{} described earlier, it is evident that malicious payloads are employed as fraudulent tactics, leading to notable distinctions between the content of on-chain transactions in comparison to benign transactions. Additionally, the diverse nature of scam techniques allows for the differentiation of each category based on the transaction content associated with specific techniques. 
Extracting key features from these distinctions will form the foundation for the detection approach outlined in Section~\ref{sec:detection}.

\noindent\fcolorbox{gray!99}{gray!10}{\parbox{0.95\linewidth}{
\textbf{Finding \#1:} \shortphish{} employs malicious payloads as fraudulent tactics, leading to notable distinctions from benign transactions. Moreover, these \shortphish{} transactions can be accurately classified into sub-categories based on various techniques utilized.
}}

\begin{table}[!t]
    \centering
    \caption{Comparison of our categorization with Etherscan phishing address nameTag.}
    \resizebox{.45\textwidth}{!}{
    \begin{threeparttable}
    \begin{tabular}{lc|lc}
    \toprule
    \multicolumn{2}{c}{ \textbf{Our Categorization}}& \multicolumn{2}{c}{Etherscan's NameTag}\\ \midrule
    \multicolumn{1}{l}{Type} & num (\#) & \multicolumn{1}{l}{Type}   & num (\#) \\ \hline
    \multicolumn{1}{l}{Ice phishing scam}&363  & \multicolumn{1}{l}{-} &   \\ \hline
    \multicolumn{1}{l}{\begin{tabular}[c]{@{}l@{}} NFT order \\scam\end{tabular}}   &80  & \multicolumn{1}{l}{-} &\\ \hline
    \multicolumn{1}{l}{\begin{tabular}[c]{@{}l@{}}Address poisoning\\ scam~\tnote{1}\end{tabular}} & 4,166& \multicolumn{1}{l}{\begin{tabular}[c]{@{}l@{}}Address poisoning\\ scam\end{tabular}}   & 3,561 \\ \hline
    \multicolumn{1}{l}{\begin{tabular}[c]{@{}l@{}}Payable function\\ scam\end{tabular}}&   15  & \multicolumn{1}{l}{-} &\\ \hline
    \multicolumn{1}{l}{Unknown}  &  506& \multicolumn{1}{l}{Unknown}& 1,569  \\ \bottomrule
    \end{tabular}
    \tnote1[*] The address poisoning label indicates a potentially harmful address that could lead to poisoning-related losses, rather than losses that have already occurred.
    \end{threeparttable}
    }
    \label{tab:label_compare}
    \vspace{-1em}
    \end{table}

\subsection{Evaluation of \shortphish{} Anatomy}
\label{subsec:evaluation_anatomy}

To ensure the coverage and effectiveness of \shortphish{} anatomy, we evaluate our classification by comparing it to well-known phishing labels. Specifically, we choose to utilize the Etherscan~\footnote{The entity information has been verified by Etherscan.} \code{Fake\_Phishing} nametags, which are the largest publicly available source of phishing nametags. However, during our investigation, we encountered certain issues with the data fetched from Etherscan.
For instance, we found that addresses belonging to the \code{Hacker} subcategory were separate and distinct from phishing and should not be included under \code{Fake\_Phishing}. Additionally, the \code{Fake token} subcategory represented simple transaction phishing, which fell outside the scope of our study. Consequently, we excluded these addresses from our analysis.
As a result, there are two effective subcategories provided by Etherscan for \shortphish{}:
\begin{itemize}
\item \textbf{Address poisoning scam.} The description states the address related to address poisoning scams, \eg, \textit{"This address may be attempting to impersonate a similar-looking address"} and \textit{"Zero Value Token Transfer Phishing"}. 
\item \textbf{Unknown.} The description lacks a specific reason, \eg, \textit{"involved with a phishing campaign"}, and \textit{"involved in suspicious activities"}.
\end{itemize}

Accordingly, we collected a total of 5,130 addresses along with their corresponding phishing labels from May 10, 2023, to July 20, 2023. The quantities of each nametag type are presented in Table~\ref{tab:label_compare}.
Comparing our classification to Etherscan, we achieved a more comprehensive coverage and broader inclusion of phishing labels. Our classification encompassed four types, providing a coverage rate of 91.2\%, with only 9.8\% of addresses labeled as \code{Unknown}. For the remaining 506 unknown addresses, we conducted additional manual analysis. Some of these labels were assigned because the addresses had been identified as phishing addresses on other EVM-compatible chains, even though they were not phishing on Ethereum. Others, according to a multi-chain search conducted by Debank~\cite{DEBANK}, were found to be completely empty addresses. Since no recorded phishing transactions were associated with these addresses on Ethereum, we were unable to classify them with the available data.

In summary, our anatomy achieves a better coverage of addresses with valid transactions, allowing for a comprehensive analysis of the phishing landscape on Ethereum.

%% file: section/02Detection.tex
\section{Detection of \shortphish{}}
\label{sec:detection}

In this section, we first introduce the key features for \shortphish{} detection based on the previous analysis. We then propose a rule-based detection approach and evaluate its effectiveness using the ground-truth dataset.

\subsection{Key features for \shortphish{} detection}
Drawing from the insights gained through the categorization (see Table~\ref{subsec:categorization}), we extract four key features for phishing transaction detection:
\begin{itemize}[leftmargin=*]
    \item[$\circ$] \textit{Contract code called by the transaction (\textbf{Code})}. For transactions involving contracts, we capture the relevant contract code, including the bytecode, .sol files, and ABI files (if the contract is open source). 
    \item[$\circ$] \textit{Transaction input data (\textbf{InputData})}. The input data of a transaction is composed of the hash of the function and its corresponding parameter arguments. We parse the input data based on the ABI file of the called contract, allowing us to extract specific function and parameter information~\footnote{In the case of phishing that abuses legitimate contracts, it is essential to note that the legitimate contracts are typically open-source.}.
    \item[$\circ$]  \textit{Transaction-related addresses (\textbf{Address})}. We collect all addresses involved in the transaction, including the caller, the callee of the transaction, and the addresses parsed from the parameter information. 
    \item[$\circ$] \textit{Transaction history (\textbf{History})}. For the $tx.from.address$ of the transaction (\ie, \code{msg.sender}), we collect their transaction history, which includes all transactions related to these addresses. 
\end{itemize}

\begin{table}[!tb]
    \caption{ \shortphish{} detection rules. The prerequisite serves as the condition criterion for a broad category (\ie, I: Ice Phishing). Upon fulfillment of prerequisites, the detection is further subdivided into respective rules based on sub-categories (\ie. I-A: Approve). }
    \label{tab:gongshitable}
    \centering
    \resizebox{.45\textwidth}{!}{
    \begin{tabular}{lll}
    \toprule
    \multicolumn{2}{c}{\textbf{Phishing Category}} & \multicolumn{1}{c}{\textbf{Rules}} \\ \midrule
    \multicolumn{1}{l}{\multirow{10}{*}{\begin{tabular}[c]{@{}l@{}}I: Ice\\Phishing\end{tabular}}}& Prerequisites& \begin{tabular}[c]{@{}l@{}}tx has transfer \&\\ tx.from $\neq$ transfer.from \&\\ tx.from $\notin$ Authorized\_List \&\\ transfer.value = transfer.from.value\end{tabular}\\ \cmidrule{2-3} 
    \multicolumn{1}{l}{}& I-A: Approve & \begin{tabular}[c]{@{}l@{}}${\exists}$ {[}'approve','increaseAllowance'{]} \\ in transfer.from.History, \\$\quad$ authorized\_address = tx.from\end{tabular} \\ \cmidrule{2-3} 
    \multicolumn{1}{l}{}& I-B: Permit& \begin{tabular}[c]{@{}l@{}}${\exists}$ {[}'permit','permit2'{]} in \\ transfer.from.History, \\$\quad$ authorized\_address = tx.from\end{tabular} \\ \cmidrule{2-3} 
    \multicolumn{1}{l}{}& I-C: setApproveForAll& \begin{tabular}[c]{@{}l@{}}${\exists}$ {[}'setApproveForAll'{]} in \\ transfer.from.History, \\$\quad$ authorized\_address = tx.from\end{tabular} \\ \midrule
    \multicolumn{1}{l}{\multirow{8}{*}{\begin{tabular}[c]{@{}l@{}}II: NFT \\ Order\end{tabular}}} & Prerequisites& tx.to.Address $\in$ NFTMarkt \\ \cmidrule{2-3} 
    \multicolumn{1}{l}{}& II-A: Bulk transfer& \begin{tabular}[c]{@{}l@{}}param.func = 'bulkTransfer' \&\\ param.recepient $\notin$ tx.from.History \end{tabular}\\ \cmidrule{2-3} 
    \multicolumn{1}{l}{}& II-B: Proxy upgrade & \begin{tabular}[c]{@{}l@{}}param = 'upgradeto' \&\\ tx.from.Address $\neq$ param.owner\end{tabular}\\ \cmidrule{2-3} 
    \multicolumn{1}{l}{}& II-C: Free buy order& \begin{tabular}[c]{@{}l@{}}param.price = 0 \\ param.fees = 100\% \\ param.recipient $\neq$ param.offerer\end{tabular} \\ \midrule
    \multicolumn{1}{l}{\multirow{4}{*}{\begin{tabular}[c]{@{}l@{}}III: Address \\ Poisoning\end{tabular}}} & Prerequisites& \begin{tabular}[c]{@{}l@{}}
    tx has transfer \& \\
    ${\exists}$ transfer' in tx.from.History.transfer \\ $\quad$ transfer'.to.address = transfer.to.address \&\\ ${\exists}$ transfer'' in tx.from.History.transfer \\ $\quad$ transfer''.to.address $\approx$ transfer'.to.address \& \\ transfer''.value \textgreater 0\end{tabular} \\ \cmidrule{2-3} 
    \multicolumn{1}{l}{}& III-A: Zero value & transfer'.value = 0 \\ \cmidrule{2-3} 
    \multicolumn{1}{l}{}& III-B: Fake token & transfer'.token $\in$ fake token \\ \cmidrule{2-3} 
    \multicolumn{1}{l}{}& III-C: Dust value & transfer'.value \textless 0.01\\ \midrule
    \multicolumn{1}{l}{\multirow{3}{*}{\begin{tabular}[c]{@{}l@{}}IV: Payable \\ Function\end{tabular}}}& Prerequisites& \begin{tabular}[c]{@{}l@{}}tx.value \textgreater 0 \&\\ close\_source(tx.to.Code) \&\\ tx.log = null\end{tabular} \\ \cmidrule{2-3} 
    \multicolumn{1}{l}{}& IV-A: Airdrop function & tx.InputData.funcsig $\in$ Airdrop \\ \cmidrule{2-3} 
    \multicolumn{1}{l}{}& IV-B: Wallet function& tx.InputData.funcsig $\in$ Wallet\\ \bottomrule
    \end{tabular}
    }
    \vspace{-1em}
    \end{table}

\subsection{Rule-based \shortphish{} detection approach}
By integrating these features, we propose a rule-based multi-dimensional detection approach. This approach involves employing customized detection methods for each category, utilizing specific detection features to ensure high accuracy. The detailed detection rules are outlined in Table~\ref{tab:gongshitable}. In the following, we elaborate on the detection methods for each phishing category:

\begin{itemize}[leftmargin=*]

\item \textbf{Ice Phishing Scam Detection}. This type of scam tactic abuses legitimate contracts.
Firstly, we collect a list of authorized addresses (called \textit{Authorized\_List}) obtained from Etherscan, including those associated with decentralized exchanges (DEX) and DeFi projects.
Next, we set up the prerequisites rule of ice phishing scam: when we encounter a transaction that involves valuable fund transfers and notice a discrepancy between the $tx.from.address$ and $transfer.from.address$, we conduct further analysis on the $tx.from.address$.
If the target address is unauthorized (not in the \textit{Authorized\_List}) and transfers all existing funds in the $transfer.from.address$, do we classify the transaction as \textit{I: Ice Phishing scam}.

To recognize each sub-category, we proceed to gather the transaction history of the $tx.from.address$ and $transfer.from.address$. Based on the various transaction types identified from the transaction history, we categorize the transaction into subcategories such as \textit{I-A: approve}, \textit{I-B: permit}, or \textit{I:C setApproveForAll}.

\item \textbf{NFT Order Scam Detection}.
This type of scam tactic abuses legitimate contracts.
First, we apply a prerequisite to isolate transactions based on the transaction callee $tx.to.address$. In this study, we only focus on the addresses that belong to the famous NFT markets (\eg, Opensea, Blur, X2Y2)~\footnote{In detail, the Seaport 1.1, Seaport 1.2, Seaport 1.3, Seaport 1.4, Blur.io Marketplace, Blur.io Marketplace 2.0, Opensea Helper, Opensea Factory}.
Then, combining the contract Code and ABI file, we parse the transaction Input Data to get the \textit{parameters}.
When the parameters meet the function \code{bulkTransfer} and the recipient is not in the transaction sendor history $tx.from.History$, we label them as the \textit{II-A: bulk transfer scam}.
Seamless, if the parameters meet the function \code{upgradeTo}, we check whether the owner is the $tx.from.address$ to judge if it is a \textit{II-B: proxy upgrade scam}.

Turn to free buy order scam, we mainly focus on the conditions given by the seller, including NFT price, receipt address, and tips.
\textit{(i)} the seller signs a sales order where the NFT price is \$0, \ie, without \code{collection} in Seaport 1.1 \code{fullfilAdvancedOrder}.
\textit{(ii)} the seller gives an incredibly high fee to the buyer, \ie, the 100\% \code{fees} in Blur \code{execute}. It results in the same result of zero buy, see Figure~\ref{fig:blur-fee}.
\textit{(iii)} the order recipient is not the NFT seller, \ie, the seller gives his \code{WETH} to the buyer in Blur \code{execute}. 
When a transaction exhibits any of these abnormal behaviors, we classify it as a \textit{II-C: free buy order scam}.

\item \textbf{Address Poisoning Scam Detection}. 
Address poisoning scams adhere to a prerequisite, regardless of the specific deceptive techniques employed (\ie, fake token, zero value, or dust transfer): 
\textit{(i)} When the victim sends a phishing transaction with a transfer (\ie, tx has transfer). The fake address already exists in historical transactions. (\ie, (\ie, ${\exists}$ transfer' in $tx.from.History$,  $transfer'.to.address$ = $transfer.to.address$))
\textit{(ii)} Before the scammer imitates a fraudulent transfer record from victim to a fake similar address, it is essential that the address has already sent valuable tokens to the genuine address, where the fake address is highly similar to the genuine address (\ie, ${\exists}$ transfer'' in $tx.from.History.transfer$, $transfer''.to.address \approx transfer'.to.address$ \& $transfer''.transfer.value$ \textgreater 0).

Specifically, when we observe that the first 4 bits and the last 4 bits of two addresses are identical, we consider these addresses to exhibit a high degree of similarity.
After we encounter a \shortphish{} transfer, we finally conduct preliminary matching of transactions with suspicious transfer behavior, \ie, zero value transfer, fake token transfer, and dust value transfer.

\item \textbf{Payable Function Scam Detection}.
The payable function scam relies on masquerading as innocuous function names to lure victims. After we observe many famous DeFi projects' functions, we observe a pattern: \textit{(i)} Most functions are open-source. \textit{(i)} Most functions are not \code{payable}, which means they can not receive users' native tokens. \textit{(iii)} Most functions have implementation logic that is not empty.

Inspired by that, we first collect function signatures with sensitive names from Ethereum 4byte Signature Database~\cite{Etherum_sig}, such as claim, claimRewards, and Claim. Based on their function name, we separate the function signatures into \textit{Airdrop} and \textit{Wallet} classes. 
Next, we establish the prerequisites for our detection approach: we consider only valuable transactions ($tx.value$ \textgreater 0) that have no associated transaction logs ($tx.log$ = null). In such cases, we attempt to retrieve the contract source code. If the source code is inaccessible (\ie., closed-source), we classify the transaction as an \textit{IV: Payable Function} scam and further classify the sub-categories (\ie, \textit{IV:A Airdrop function}, \textit{IV:B Wallet function}) based on the corresponding function signatures.

\end{itemize}

\subsection{Evaluation of \shortphish{} detection approach} 
\label{subsec:evaluation}
We have implemented a prototype to evaluate our detection approach. First, to expedite the collection of Ethereum transaction information, we set up a local Ethereum archive node following the methodology described by Feng et al.~\cite{feng2024slimarchive}. Additionally, to speed up the history data collection, we accelerated the historical transaction replay process by following the techniques outlined by Wu et al.~\cite{wu2022time}. Finally, we implemented our aforementioned detection rules using Golang.

Besides the prototype implementation, we collected two datasets, \ie the ground-truth dataset (see Section~\ref{subsec:informaton_collect}), and a large-scale dataset consisting of Ethereum transactions from May 1, 2023, to Jun 1, 2023. The large-scale dataset includes 210,000 blocks with 30,976,209 transactions.
In the following sub-sections, we will first use the ground-truth dataset to assess the accuracy of our approach. After that, we will apply our approach to the large-scale dataset to evaluate its real-world accuracy and efficiency.

\subsubsection{Accuracy Evaluation}

\renewcommand{\DTstyle}{\textrm}
\begin{table}[!t]
\caption{Accuracy evaluation of the detection approach.}
\label{tab:ground-truth-data}
\vspace{-1em}
\center{
\begin{threeparttable}
\begin{tabular}{lrr}
\toprule
\textbf{Category} & \textbf{ Num (\#)} & \textbf{TP/FP/FN}\\
\midrule
\DTsetlength{0.2em}{0.7em}{0.2em}{0.4pt}{0pt}
\begin{minipage}{2.8cm}\dirtree{%
.1 \textbf{Ground-truth}.
.2 \textbf{Benign}.
.2 \textbf{\shortphish{}}.
.3 Ice phishing.
.3 NFT order.
.3 Address poisoning.
.3 Payable function.
.1 \textbf{Large-scale}.
.2 \textbf{\shortphish{}}.
}\end{minipage}
&
\DTsetlength{0pt}{0pt}{0pt}{0pt}{0pt}
\begin{minipage}{1cm}\dirtree{%
.1 .
.1 \textbf{13557}.
.1 \textbf{5000}.
.1 2569.
.1 609.
.1 226.
.1 1596.
.1 .
.1 -.
}\end{minipage}
&
\DTsetlength{0pt}{0pt}{0pt}{0pt}{0pt}
\begin{minipage}{1cm}\dirtree{%
.1 .
.1 \textbf{13555/1/2}.
.1 \textbf{4999/2/1}.
.1 2568/0/1.
.1 609/0/0.
.1 226/0/0.
.1 1596/2/0.
.1 .
.1 \textbf{12050/84/6}~\tnote1.
}\end{minipage}
\\
\bottomrule
\end{tabular}
\begin{tablenotes}
\footnotesize
\item[1] For false negatives, we manually reviewed not detected as PTXPHISH transactions but initiated by addresses labeled as  \texttt{Fake\_Phishing} by Etherscan within the same timeframe.
\end{tablenotes}
\end{threeparttable}
}
\vspace{-1em}
\end{table}

For the accuracy evaluation on the ground-truth dataset, we conducted separate accuracy assessments for each phishing category, as shown in the table~\ref{tab:ground-truth-data}. 
The table demonstrates that our detection approach achieves remarkably high accuracy on the ground-truth dataset, with an overall F1-score over 99.9\% (only 2 FPs in payable function and 1 FN in ice phishing).

For the large-scale dataset, we detected 12,050 \shortphish{} transactions. To evaluate false positives (FPs), our research team manually reviewed these transactions using the process described in Section~\ref{subsec:informaton_collect}. 
However, manually evaluating false negatives (FNs) in the same manner was impractical due to the large volume of transactions.
Therefore, for transactions not detected as \shortphish{}, we collected their initiating addresses to check if they were flagged as \code{Fake\_Phishing} by Etherscan within the same timeframe. We then manually reviewed transactions initiated by addresses labeled as \code{Fake\_Phishing}. If a transaction was confirmed to be phishing, it was classified as an FN.
The results are summarized in Table~\ref{tab:ground-truth-data}: 84 transactions were identified as FP (4 in ice phishing and 80 in misleading), and 6 transactions were identified as FN (all in NFT order), resulting in an overall F1-score of 99.6\%.

Additionally, we conducted a manual analysis to clarify instances of false detection cases.
For ice phishing, the majority of FPs resulted from victims approving transactions to themselves and invoking the \code{transferFrom} function. This rare behavior closely mimicked phishing activities and could not be distinguished by our detection approach.
FPs related to the payable function were attributed to specialized Miner Extractable Value (MEV) bots that employed payable functions without logical functionalities. 
These MEV bots had off-chain information beyond our knowledge, leading to FP occurrences.

Regarding FNs, most were observed in ice phishing and NFT orders. 
In ice phishing, FNs resulted from scammers leveraging decentralized exchanges (DEX) to convert victims' funds into alternative tokens, with the phishing address as the recipient. The complex contract semantics of these swaps disrupted the flow of funds, leading to FNs.
In NFT orders, FNs occurred because some scammers used extremely low prices (\eg, 1 wei) to perform free order tricks instead of 0 value, resulting in detection failures.

These special cases will be discussed further in Section~\ref{sec:dis}.

\begin{table}[]
    \centering
    \caption{Efficiency evaluation of the detection approach. T/B means the time consumption per block.}
    \resizebox{.45\textwidth}{!}{%
    \small
    \begin{tabular}{c|ccc}
    \toprule
    Ethereum        & \multicolumn{3}{c}{Detection Approach}             \\ \hline
    Ave.T/B & \multicolumn{1}{c|}{Ave. T/B} & \multicolumn{1}{c|}{Median T/B} & Max T/B\\ \hline
    12,000 ms            & \multicolumn{1}{c|}{390 ms}  & \multicolumn{1}{c|}{362 ms}         & 3,553 ms         \\ \bottomrule
    \end{tabular}
    }
    \label{tab:time}
    \vspace{-1em}
    \end{table}

\subsubsection{Efficiency Evaluation}
To evaluate the efficiency of our detection approach, we use real Ethereum blocks to calculate time consumption. In Ethereum, the fundamental unit of packaging is the block, which contains multiple transactions. The average block production time in Ethereum is 12 seconds (12,000 ms). 
As shown in Table\ref{tab:time}, our approach exhibits high efficiency, with an average time consumption of just 390 ms per block, a median of 362 ms per block, and a maximum of 3,553 ms. 

For a more detailed view, we present a time consumption graph in Figure~\ref{fig:timeconsumption} in the appendix. 
Our approach consistently consumes significantly less time than the block production time, even for blocks with the maximum time (which are rare, making the average time consumption a more reliable metric). 
Therefore, our approach meets the requirements for real-time performance and has been integrated into Forta, a well-known real-time anti-phishing platform (detailed in Section~\ref{sec:dis}).

%% file: section/03transaction_ana.tex
\section{Large scale detection in the real world}
\label{sec:ana}
Given the demonstrated effectiveness of the proposed detection approach in the previous section, we can now apply this approach to detect real-world threats. Specifically, we conduct the detection on the Ethereum blockchain, covering the period from block \textit{16,304,348} to block \textit{18,440,040}. This corresponds to a timeframe of 300 days, spanning from December 31, 2022, to October 27, 2023. During this period, our detection approach identifies a total of \textbf{130,637} \shortphish{} transactions, as detailed in Table~\ref{tab:collection_phish_new}.

Building upon the detection results, we proceed to perform a comprehensive analysis in various aspects. In Section~\ref{subsec:txanalysis}, we delve into an analysis of the \shortphish{} transactions themselves. Section~\ref{subsec:scam} focuses on examining the characteristics and behaviors of \shortphish{} scammers, while Section~\ref{subsec:vic} explores the experiences and impact on victims of such scams. Lastly, in Section~\ref{subsec:contribution}, we present the valuable action we have provided to help combat and mitigate the risks posed by these real-world threats.

\subsection{Analyzing \shortphish{} Transactions}
\label{subsec:txanalysis}
To analyze the \shortphish{} transactions, we present our analysis from multiple perspectives. First, We examine the economic losses caused by phishing and their relationship with time changes. Secondly, we analyze the characteristics and performance of different phishing categories.

\noindent \textbf{The economic losses caused by \shortphish{}.}
Considering the diversity of asset types and price fluctuations, it is important to explain the principles for calculating losses. 
To ensure that our calculations are as realistic as possible, the prices of all ERC-20 tokens and NFTs are chosen as the price when the phishing transaction occurs. 
Specifically, the price of ERC20 is determined by the price oracle~\footnote{In this study, we only focus on Top tokens, \ie, ETH, USDT, USDC, DAI, WETH, stETH, WBTC, BUSD.}. To NFTs, there is currently no way to determine the price of a specific NFT, we use the floor price marked on OpenSea instead. 

\begin{table}[]
    \centering
    \caption{Detected \shortphish{} and losses in 300 days.}
    \resizebox{.45\textwidth}{!}{%
    \small
    \begin{threeparttable}
    \begin{tabular}{lccc}
    \toprule
    \multicolumn{1}{c}{\textbf{Phishing Category}} & \multicolumn{1}{c}{\textbf{Number (\#)}} & \textbf{Loss (\$)}   & \textbf{Average (\$)} \\ \midrule
    Ice phishing& 47,762   & 201,880,314 & 4,226.8\\ \midrule
    NFT order   & 14,999   & 57,495,168  & 3,833.3\\ \midrule
    Address poisoning   & 1,050& 64,042,825  & 60,993.2  \\ \midrule
    Payable Function& 66,826   & 18,527,500  & 277.3 \\ \midrule
    \textbf{Total}   & \textbf{130,637}  & \textbf{341,945,807} & \textbf{2,617.5}\\ \bottomrule
    \end{tabular}
    \end{threeparttable}
    }
    \label{tab:collection_phish_new}
    \vspace{-1em}
    \end{table}

We summarize the detailed phishing transactions and corresponding losses based on their phishing category in Table~\ref{tab:collection_phish_new}.  
In total, \shortphish{} caused a total loss of \$341,945,807 during 300 days.
Among them, ice phishing has the highest proportion, accounting for \$201,880,314 (59.04\%). 
Address poisoning is the second highest, with a total profit of \$64,042,825 accounting for 18.73\%. 
Market order scams generate a total profit of \$57,495,168, accounting for 16.81\%. Finally, payable function scams generate \$18,527,500, accounting for 5.42\%.  Interestingly, when we calculate the average losses, we observe variations in the profit strategies employed by phishing scams. For example, the number of address poisoning scams is relatively small (only 1,050 cases), yet they yield a significant individual loss of \$60,993 per transaction. In contrast, payable function scams have the highest occurrence rate (66,826 cases), but the individual transaction loss is only \$273.

\begin{figure}[!t]
\centering
\includegraphics[width=.45\textwidth]{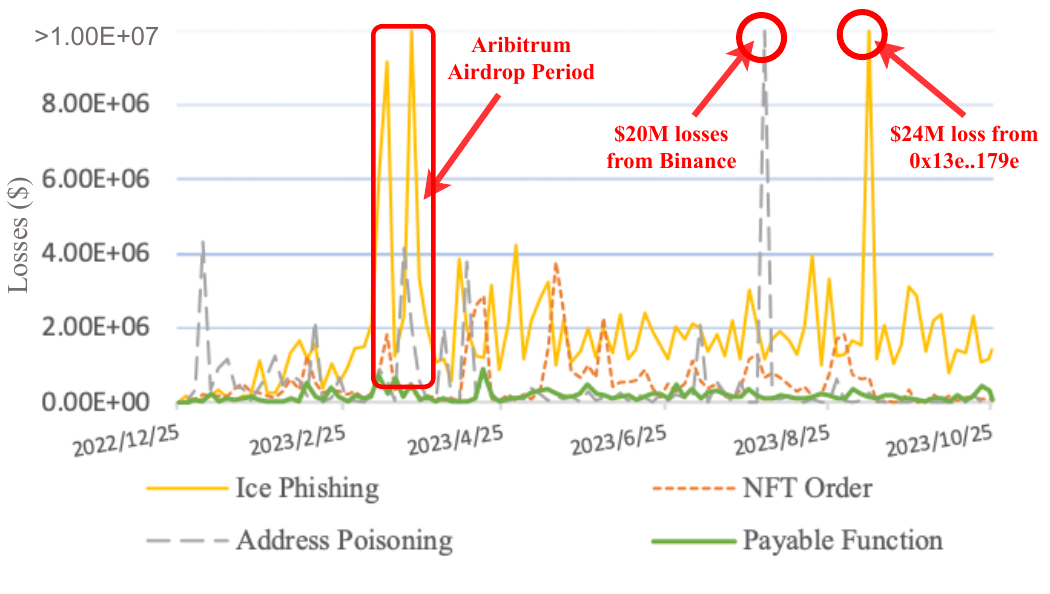}
    \caption{Variation of \shortphish{} losses over time.}
	\label{fig:timeloss}
    \vspace{-1.5em}
\end{figure}

We conclude the graph of \shortphish{} by data and corresponding losses in Figure~\ref{fig:timeloss}, from which we can see that \shortphish{} has existed for a long time since early 2023, without being effectively solved, and as time goes by, the losses are still increasing.
It can be seen that phishing is an increasingly and continuously serious social problem, which further highlights the value of our work.
Especially, from March 22 to 24, the losses amount reached over \$30 million.
After investigating the dates of these extreme cases, we find that Arbitrum airdrops~\cite{Arbair} occurred on March 23, 2023. Unfortunately, such campaigns often result in great phishing success. 
In the later stage, we find two extremely high losses, \ie, \$20M from the address poisoning attack suffered by Binance and \$2.4M losses from the ice phishing of victim 0x13e382dfe53207E9ce2eeEab330F69da2794179E.
To examine the evolution and emergence process of elaborate scams, we conducted a separate study on the active periods of various scams in the early months of 2023, as illustrated in Figure~\ref{fig:heatmap} in the appendix.
From the active period of different phishing sub-categories, it is evident that these phishing methods are constantly evolving and improving. 
For instance, zero value transfer poisoning was already been active on December 25, 2022, as an early phishing method. 
However, with the emergence of new variants of address poisoning scams, the first successful phishing transaction of dust poisoning appeared on March 7, 2023, while the first successful fake token poisoning appeared on March 16, 2023. The time interval shows that scammers continue to innovate fishing methods.

\noindent\fcolorbox{gray!99}{gray!10}{\parbox{0.95\linewidth}{
\textbf{Finding \#2:} 
\shortphish{} has become a threatening cybercrime, yielding profits exceeding \$341.9 million during a 300-day observation period. To make the most profit, \shortphish{} employ different strategies. Payable function scams are numerous with small profits per transaction. In contrast, address poisoning scams are fewer in number but can generate significant profits in a single instance.
}}
\vspace{1em}

\noindent \textbf{The characteristics of each \shortphish{} category.}
To better understand the characteristics of \shortphish{} tricks, we delve into each trick respectively. 
\begin{itemize}[leftmargin=*]
    \item  \textbf{Ice Phishing Scam.} To better understand the prevalence of ice phishing scams, we conduct an analysis of the total number of \code{approve} and \code{permit} transactions during the same period.
    Our findings reveal that, out of the token contracts we examined, there are a total of 4,207,423 successful \code{approve} transactions. Among these, 209,318 transactions are identified as phishing approves, accounting for 4.97\% of the total number.
    Even more concerning, we discover that out of the 13,877 successful \code{permit} transactions, 6,414 transactions are identified as phishing permits, accounting for a staggering 46.22\% of the total number.
    These alarming numbers show that the \code{approve} and \code{permit} functions are abused by phishing scams. 
    Based on our speculation, these functions are favored by phishers due to their hidden and efficient ability to transfer funds ownership.
    
    \item \textbf{NFT Order Scam.}
    We analyze the movement of stolen NFT assets.
    In total, there are 61,838 stolen NFTs, of which 16,442 NFTs have been transferred after being stolen (only 26.6\%) until November 30, 2023. It indicates the poor liquidation of NFT assets.
    In addition, we tracked the transfer events of these NFTs and identified the NFT markets in which these NFTs are sold. Finally, the movements of stolen NFTs are summarized in Table~\ref{tab:nft_market} in the appendix. 
    From the table, we find that most scammers (62.22\%) directly sell the NFTs to the market using the cashier address, while a small portion (17.85\%) transfers NFTs to fund aggregators for selling.
    Among the stolen NFTs, we observed that most of the NFTs were sold through Blur (61.78\%), followed by OpenSea (21.97\%), X2Y2 (8.32\%), and LooksRare (7.83\%).
    In summary, we conclude that these NFT marketplaces do not effectively prevent the sale of stolen NFTs, and over 80\% of stolen NFTs are sold through the markets.

\item \textbf{Address Poison Scam.}
To poison the victims' transaction history, scammers will actively initiate attack transactions.
During the observation phase, we discovered a total of 888,744 address poisoning attack transactions, resulting in a total of 3,132,607 addresses being affected.
This long-term and extensive scam method poses a significant threat to the security of all addresses.

The scammer needs to pay the gas fee for their attack transactions. We calculate and find that the gas fee consumed by the scammers was 4023.3 ETH over a period of 300 days (13.4 ETH daily).  Additionally, \$60,509 tokens were used for dust transfers. According to Etherscan~\cite{Etherscan}, the daily gas consumption is around 107.5 ETH, which means that the gas fee consumed by address poisoning attack transactions accounts for 12.5\% of all gas fees on the entire Ethereum.

\item \textbf{Payable Function Scam.} 
We conduct an analysis of various functions used in payable function scams to determine their respective proportions (see Table~\ref{tab:contract} in the appendix). The total loss resulting from these scams exceeds \$18 million. We observe two distinct types based on their functionalities: \textit{Airdrop} accounted for 74.2\% of the total losses (e.g., \code{Claim}/\code{claim}), while \textit{Wallet} accounted for 25.8\% (e.g., \code{SecurityUpdate}).
These findings indicate that victims of this specific phishing attack are primarily motivated by greed, as they aim to profit from potential gains associated with accepting airdrops. Unfortunately, their funds are ultimately stolen through deceptive profit-generating mechanisms employed by scammers. It is crucial to note that a minority of victims lack a fundamental understanding of blockchain technology and mistakenly perceive these interactions as standard wallet operations. As a result, they unknowingly make payments and become prey to these phishing scams.
\end{itemize}

\noindent\fcolorbox{gray!99}{gray!10}{\parbox{0.95\linewidth}{
    \textbf{Finding \#3:} \shortphish{} is extremely rampant and has impacted ecosystem of Ethereum. 
    For example, 4.97\% \code{approve} transactions and  46.22\% \code{permit} transactions are identified as phishing transactions.
    Scammers consume about 4023.3 ETH as transaction fees (13.4 ETH daily) to spread the address poisoning scams, which account for 12.5\% of the total Ethereum gas fees.
    }}

%% file: section/04attacker_ana.tex
\subsection{Analyzing \shortphish{} Scammer}
\label{subsec:scam}

\begin{figure}[!t]
\centering
	\includegraphics[width=.4\textwidth]{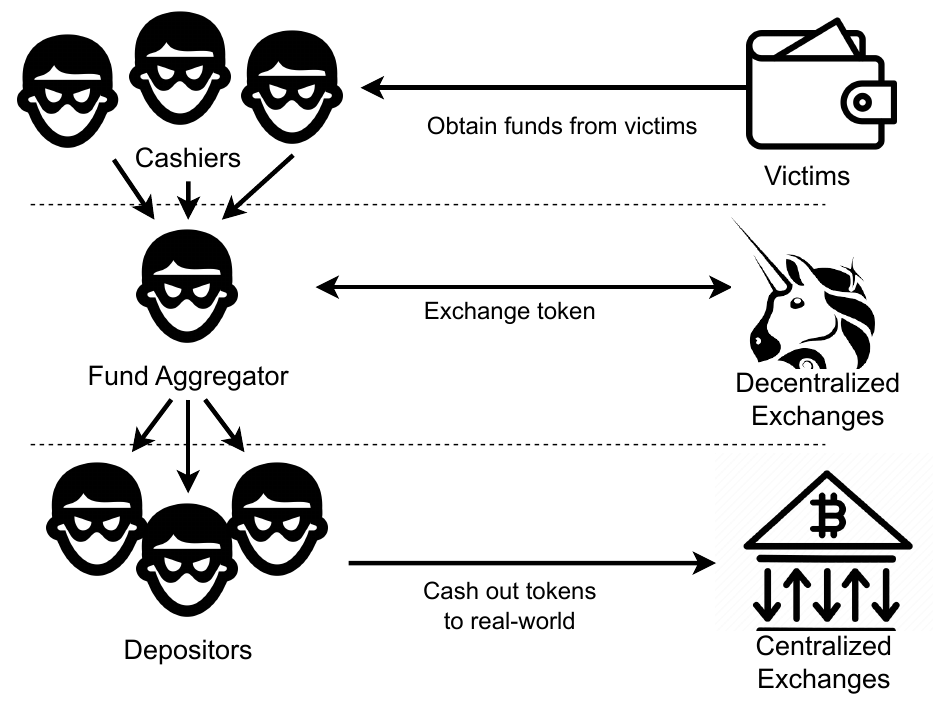}
    \caption{Scammer organization during the cash-out process.}
	\label{fig:family}
    \vspace{-1.5em}
\end{figure}

In this section, we analyze the \shortphish{} scammer and focus on their fund flow during the cash-out process, \ie, the money transfer pattern and scammer address organization.  
After reviewing scams that occurred over six months, we find a special money cash-out pattern, and categorize the behavior of scammer addresses into the following three types:
\begin{itemize}[leftmargin=*]
    \item[$\circ$] \textit{Cashiers.} The Cashier addresses are responsible for directly obtaining funds from victims.
    \item[$\circ$] \textit{Fund Aggregators.} 
    The fund aggregator addresses are responsible for aggregating the profit funds from multiple cashier addresses~\footnote{During our analysis, we found that fund aggregators always receive funds from more than 3 cashier addresses.}. 
    The fund aggregators may also be involved with multiple DeFi protocols, such as token swaps in decentralized exchanges (DEXes).
    \item[$\circ$] \textit{Depositors.} The depositor addresses are responsible for depositing on-chain assets to centralized exchanges (CEXes).
\end{itemize}

We illustrate the money cash-out pattern and the address organization in Figure~\ref{fig:family}. 
First, the cashier addresses obtain funds from victims. Then, multiple cashiers transfer their funds to the fund aggregator.
The fund aggregator may exchange the tokens into fiat currencies (\eg, USDT, USDC) or Ether.
Finally, the fund aggregator transfers the funds to multiple depositor addresses, which cash out the profits by CEXes. 
Additionally, to escape the regulation from CEX and security companies, the fund aggregator addresses will change occasionally, resulting in a cashier address transferring money to different fund aggregators.
Due to the complex DeFi semantics and large transaction volume (over 3 billion transfers until June, 2023), the money flow graphs (MFG) of blockchain are complex and over-weight for analysis~\cite{kalodner2020blocksci,wu2021towards}. 
However, based on the cash-out pattern, we propose a lightweight organization discovery algorithm based on their fund flow relationships and scammer roles, and show the algorithm process as follows.
\begin{itemize}[leftmargin=*]
    \item[$\circ$] \textit{Step S1: Locate the Cashier.} 
    First, we collected a set of cashier addresses from \shortphish{} transactions that were directly exposed and identified as recipients of stolen funds. 
    \item[$\circ$]  \textit{Step S2: outgoing Transfer Expansion.} 
    We trace the outgoing fund transfer of the cashier addresses and record the destination addresses. To make the outgoing fund transfer more reliable, we analyze several famous DEXes (\eg, Uniswap, Sushiswap), and remove redundant edges caused by DEX interaction. What's more, we prune transfers with a small value (less than \$100). 
    \item[$\circ$]  \textit{Step S3: Expansion Address Categorization.}
    After getting the outgoing transfer destination addresses, we further categorize these addresses through behavior features: 
    (i) if the destination address is in the CEX whitelist (the CEX whitelist is collected from Etherscan), the address is labeled as the CEX address. 
    (ii) if there are over 3 cashiers with the same outgoing destination address, we label the destination address as the fund aggregator address; 
    (iii) if the address does not fall into either of the above categories, we label it as an unknown address.
    \item[$\circ$] \textit{Step S4: Repeat Expansion \& Categorization.} For the remaining unknown addresses, we further trace their outgoing transfers like step S2. And perform address categorizations like step S3. In this study, we repeat 3 times in total. 
\end{itemize}

\begin{table}[!tb]
    \caption{Top 5 scammer organizations and profits.}
    \centering
    \resizebox{.45\textwidth}{!}{
    \begin{tabular}{llcc}
    \toprule
    Rank & Famous Address & Total Profits (\$) & Percentage \\ \midrule
    \#1 & - &     60,149,219 & 17.6\%  \\ \midrule
    \#2 & \multirow{2}{*}{\begin{tabular}[c]{@{}l@{}}Fake\_Phishing186943 \\ Fake\_Phishing186944~\cite{Largest_phish_appr}\end{tabular}} &     31,628,798 & 9.25\%  \\ \\\midrule
    \#3 & \multirow{2}{*}{\begin{tabular}[c]{@{}l@{}}Fake\_Phishing179050 \\(Sha zhu pan~\cite{shazhupan})\end{tabular}} & 17,854,190 & 5.2\%\\ \\ \midrule
    \#4 & VenomDrainer~\cite{Drainer} &   16,351,857 & 4.8\%  \\ \midrule
    \#5 & \multirow{2}{*}{\begin{tabular}[c]{@{}l@{}}InfernoDrainer~\cite{infernoDrainer} \\ AngelDrainer~\cite{Angel_drainer}\end{tabular}}  &  \multirow{2}{*}{13,196,846} & \multirow{2}{*}{3.9\%} \\  \\   \bottomrule
    \end{tabular}
    }
    \label{tab:family_sample}
    \end{table}

We show our algorithm process in Figure~\ref{fig:familyturn} in the appendix due to the page limit. 
In total, from the detected \shortphish{} transaction, we identified 121 scammer organizations with the same fund aggregators.
We show the top 5 scammer organization in Table~\ref{tab:family_sample}.
From the table, we can observe that among the highest-ranked organizations, there are several well-known scam addresses (\ie, \code{Fake\_Phishing186944}~\cite{Largest_phish_appr}, \code{Fake\_Phishing179050}~\cite{shazhupan}) and scam drainers (\ie, VenomDrainer~\cite{Drainer}, InfernoDrainer~\cite{infernoDrainer}, and AngelDrainer~\cite{Angel_drainer}) exposed by the media. These top organizations account for 40.7\% of all phishing scam revenue, making them a serious problem that needs to be addressed.

The findings from our cash-out pattern analysis indicate that our proposed scam organization is widely adopted within the current landscape of on-chain scam organizations. 
Nevertheless, our proposed pattern has certain limitations when it comes to centralized services such as underground money laundering service, which will lead to some false correlations. However, our proposed cash-out pattern can serve as an inspiration for future research endeavors that aim to uncover more fraudulent addresses by exploiting address correlations.

\noindent\fcolorbox{gray!99}{gray!10}{\parbox{0.95\linewidth}{
\textbf{Finding \#4:} Phishing addresses are highly organized during the cash-out process, with different roles such as cashier, fund aggregator, and depositor. Based on the cash-out pattern, we find that the top five phishing organizations account for 40.7\% \shortphish{} losses.
}}

\subsection{Analyzing \shortphish{} Victims}
\label{subsec:vic}

In this section, we analyze the phishing victims. 
Specifically, we conduct research on the victim's behavior profile and remedial measures after being phished.  

\noindent \textbf{The victim behavior profile.} 
Aim to identify user characteristics that are vulnerable to phishing scams.
We collected victim addresses from all phishing transactions and recorded the transactions actively initiated by these addresses.
To better demonstrate the behavior of victim addresses, we present two dimensions in Figure~\ref{fig:txloss} in the appendix, \ie, the victims' transaction volumes and corresponding losses, and the proportion of transaction types.
From the analysis of the figure, it is evident that the majority of victims have fewer than 1,000 transactions. Interestingly, in incidents involving large amounts (over \$100k), victim transactions are predominantly concentrated at less than 50. This statistic implies that experienced users with higher transaction amounts exhibit a greater awareness of phishing prevention.
Furthermore, our findings reveal that 99\% of the victims had been engaged in DeFi activities, with 20\% of them specifically involved in NFT transactions. In contrast, only 1\% of the victims were found to be engaged in simple Ethereum transfers. This data further solidifies the notion that this new phishing technique predominantly targets DeFi users.

\noindent \textbf{The victim remedial measure.}
We mainly focus on the victims of ice phishing, as this type of fraud has ongoing harm until the victim uses the \code{revoke} function to cancel the phishing approval.
According to our observations, after being ice phished, victims mainly exhibit the following three behaviors: \textit{(i)} revoking the phishing approval; (\textit{(ii)}) transferring all assets to other addresses and abandoning the victim address; \textit{(iii)} taking no remedial measures.
Out of the randomly selected 5,000 victims (in table~\ref{tab:revoke} in the appendix), only 1,316 addresses (26.32\%) chose to revoke the phishing approval, while 1,665 addresses (33.3\%) transferred all funds to other addresses, abandoning the previous address.
However, a concerning 2,019 addresses (40.38\%) did not take any remedial measures, leaving them vulnerable to further attacks and potential financial loss. 
This indicates that many victims have no idea how to take remedial measures. 
The vast majority of victims (73.68\%) did not take the most effective measure of revoking the phishing approval, but instead chose to transfer funds, which is more time-consuming and expensive. 

\noindent\fcolorbox{gray!99}{gray!10}{\parbox{0.95\linewidth}{
\textbf{Finding \#5:} The majority of victims (99\%) are actively involved in DeFi, including NFT transactions. However, a significant portion of these victims (40.38\%) lack awareness of the necessary steps to take for implementing remedial measures after experiencing a phishing attack.
}}

%% file: section/05reality_contribution.tex
\subsection{Contributing to the Community}
\label{subsec:contribution}
To further assist users in mitigating threats, we actively contribute to the community by submitting identified phishing addresses to \textit{Etherscan}, which is the largest and de-facto standard blockchain explorer on Ethereum. It offers a \textit{nametag} mechanism that allows trustworthy third parties to label various types of addresses. This practice is widely adopted by the community, including security companies and community sleuths, to combat phishing scams.
During the period from December 31, 2022, to October 27, 2023, we contributed a total of \textbf{1,726} phishing addresses. Among all the community contributors, our phishing address labels~\footnote{Etherscan only records the label source of the first submission.} accounted for \textbf{42.7\%} of the total, as shown in Table~\ref{tab:label_source}.

\begin{table}[]
    \centering
    \caption{Overview of community contributed phishing addresses.}
    \resizebox{\linewidth}{!}{
    \begin{threeparttable}
    \begin{tabular}{llll}
    \toprule
    Label Source   & \textbf{Our Reports}          & Blockmage~\cite{Blockmage} & Tayvano~\cite{Tay}      \\ \midrule
    Number (\#/\%) & \textbf{1726 (42.7\%)} & 559 (13.8\%)             & 530 (13.1\%) \\ \midrule
    Label Source   & AnciliaInc~\cite{Ancilia}    & ZachXBT~\cite{Zach}                  & Others       \\ \midrule
    Number (\#/\%) & 499 (12.3\%)  & 416 (10.3\%)             & 314 (7.8\%)  \\ \bottomrule
    \end{tabular}
    \end{threeparttable}
    }
    \vspace{-1em}
    \label{tab:label_source}
    \end{table}

In addition to providing phishing address labels to the community, we have made other efforts to assist users. Firstly, we proactively send on-chain messages directly to victims to alert them about phishing attempts. Our process involves monitoring the Ethereum pending pool for any suspicious transactions. Upon identifying a phishing transaction in the pending pool, we promptly send a transaction to the victim containing alert information.
By receiving our alert transactions, victims are empowered to take proactive measures and prevent phishing losses. During the specified period, we have successfully sent a total of \textbf{2,539} on-chain alert messages, providing assistance to \textbf{1,980} victims. Additionally, we contribute to anti-phishing efforts by providing phishing reports as online educational resources. These reports have been visited by a significant number of users, with a total visit count of \textbf{18,585} based on our internal records for that period. This effectively raises awareness and promotes anti-phishing initiatives.

As a result of our efforts, we have received expressions of gratitude in the form of on-chain transactions and tweets~\footnote{We can provide them if needed for review purposes.}. We take pride in the acknowledgment and appreciation we have received from Etherscan and other members of the community. Their recognition validates our commitment to combatting phishing attempts and protecting individuals from these threats.

%% file: section/06Discussion.tex
\section{Discussion}
\label{sec:dis}
Our study performs the first empirical study of \shortphish{}.
Although our focus is primarily on Ethereum, our approach can be easily applied to other EVM-compatible blockchains (\eg, BNB smart chain and Polygon Mainnet).
In the following, we will discuss details related to the anatomy, corner cases, and anti-phishing tools/platforms.

\noindent \textbf{Anatomy of \shortphish{}.}
In this study, we categorize the current phishing scams into four categories. However, as discussed in Section~\ref{subsec:txanalysis}, scammers are continuously developing new methods. Therefore, the categorization presented in this study reflects the current state of phishing techniques. Future advancements in phishing methods may necessitate adjustments to this categorization.

\noindent \textbf{Corner Case of Detection.} In some extreme theoretical scenarios, our detection approach may produce inaccurate results, such as self-approvals or closed-source MEV bot (see Section~\ref{subsec:evaluation}). 
Other potential corner cases might include situations where a drainer executes a \code{transferFrom} but leaves some funds with the victim. 

These cases are counter-intuitive, as we assume that all on-chain behaviors are driven by rational actors seeking to maximize their benefits.
However, behaviors like self-approvals or leaving funds behind lead to unnecessary losses or wasted gas fees, making them relatively rare. 
Consequently, while our detection approach may not cover all extreme theoretical cases, it remains suitable and effective for real-world applications.

\noindent \textbf{Anti-Phishing Tools/Platforms.}
Many security companies have developed anti-phishing tools/platforms to combat the prevalence of phishing scams. 
We list prominent anti-phishing tools/platforms in Table~\ref{tab:detect_tool} in the appendix, and categorize them based on their approaches.
Current anti-phishing tools (\eg, AegisWeb3, Pocket Universe) primarily use transaction pre-execution to predict fund changes and implement blacklists for receiving address detection. 
In contrast, our detection approach adopts a rule-based strategy based on on-chain information.
This unique approach complements existing tools and enhances their security coverage. 
Indeed, \textbf{our detection approach has been integrated into Forta}, a leading scam detection platform, establishing us as a primary partner.

%% file: section/07Relatedwork.tex
\section{Related Work}
\subsection{Security Issues on Ethereum}
Since its inception, Ethereum has faced numerous security issues. The security issues have evolved with the development of the platform. The academic community has shown great concern for the security of Ethereum, with many research~\cite{chen2018detecting,hu2022scsguard,li2022ttagn} efforts dedicated to addressing its security challenges.
Xia et al.~\cite{xia2021trade} perform the first analysis on the fake ERC-20 tokens, and leverage AI to perform fake token detection.
Chen et al.~\cite{chen2020finding} conduct analysis on smart contracts and propose a method to find the security issues by comparing historical versions.
Liu et al.~\cite{liu2022finding} focus on the permission bugs in the DeFi project, and propose a prototype detection system.
Su et al.~\cite{su2021evil} measure the DeFi attacks and propose a detection algorithm. 
Das et al.~\cite{das2022understanding} perform an in-depth analysis of the NFT ecosystem, and raise several security issues. 

\subsection{Phishing Analysis}
Research into analyzing phishing behaviors have been evolving for years. 
For traditional Web2 phishing, several studies~\cite{oest2020sunrise,lain2022phishing,lin2021phishpedia,oest2020phishtime} have analyzed phishing behaviors and characteristics.
Web3 phishing, while similar to traditional Web2 phishing, extends beyond websites and leverages cryptocurrency as a payment method~\cite{li2023double}.
He et al.~\cite{he2023txphishscope} and Li et al.~\cite{li2023double} have proposed website-based phishing detection systems and conducted analyses of phishing websites.

In addition to traditional phishing scams, Ivanov et al.~\cite{ivanov2021targeting} were the first to highlight scams exploiting misleading EVM features, \ie, address manipulation and Unicode attacks. Ye et al.~\cite{ye2023revealing} focused on phishing that involves misleading information on the wallet UI (including token symbols, wallet addresses, and smart contract function names), though their study was limited to zero-value transfers and fake \code{claim} functions.
Kim et al.~\cite{kim2023drainclog} focused on NFT scams and developed a detection model using features like price differences, time duration, and transfer relations. Li et al.~\cite{li2021measuring} collected illicit addresses from the Blockchain Intelligence Group and employed machine learning techniques to predict these addresses.

Our study distinguishes itself from related research in the following aspects: \textit{(i) Different target \& motivation.} To our knowledge, our study is the first to provide a comprehensive analysis of \shortphish{}. We aim to thoroughly investigate this new form of phishing, which may include subclasses of previous phishing tactics such as ``setApproveForAll'' in NFT phishing and ``zero value transfer'' in address poisoning.
\textit{(ii) Different detection method \& capability.} Previous research primarily relies on past fund flows, which may overlook/delay the detection of newly created phishing addresses.
In contrast, our rule-based detection method allows for \textbf{real-time} identification of phishing transactions and addresses.

%% file: section/08Conclusion.tex
\section{Conclusion}
\label{sec:conclusion}

This paper presents the first comprehensive study of \shortphish{} on the Ethereum.
First, we conducted a long-term data collection to establish the first ground-truth \shortphish{} dataset consisting of 5,000 phishing transactions. Then we dissected \shortphish{}, categorizing phishing tactics into four primary categories and eleven sub-categories.
Second, we proposed a rule-based multi-dimensional detection approach to identify phishing transactions, achieving over 99\% F1-score.
Finally, we conducted an in-depth analysis of the large-scale detection results to offer insightful findings. 
Our analysis revealed that \shortphish{} resulted in losses exceeding \$341.9 million within a 300-day period. 
Scammers expended approximately 13.4 ETH daily, which accounted for 12.5\% of the total Ethereum gas fees, in spreading address poisoning scams. Notably, the top five phishing organizations were responsible for 40.7\% of the total losses.
Furthermore, our work made significant contributions to the community. We reported a total of 1,726 phishing addresses, accounting for 42.7\% of the total community contributions during the same period. Additionally, we sent 2,539 on-chain alert messages, providing assistance to 1,980 victims of phishing attacks.